\documentclass[aps,twocolumn,prl,preprintnumbers,showpacs,amsmath,amssymb,superscriptaddress]{revtex4-1}
\usepackage[pdftex, colorlinks, citecolor=blue]{hyperref}   
\usepackage{graphicx}
\usepackage{dcolumn}
\usepackage{threeparttable}
\usepackage{multirow}
\usepackage{booktabs}
\usepackage{txfonts}
\usepackage{xcolor}
\usepackage{bm}
\usepackage{amssymb}
\usepackage{amsmath}
\usepackage{latexsym}
\usepackage{epsfig}
\usepackage{amsbsy}
\usepackage{array}
\usepackage{tabularx}
\usepackage{esvect} 
\usepackage{extarrows}
\usepackage{float}
\usepackage[british]{babel}

\begin{document}

\title{Combining Machine Learning and Many-body Calculations: \\
Coverage-Dependent Adsorption of CO on Rh(111)}

\author{Peitao Liu}
\email{ptliu@imr.ac.cn}
\affiliation{University of Vienna, Faculty of Physics and Center for Computational Materials Science, Kolingasse 14-16, A-1090 Vienna, Austria}
\affiliation{Shenyang National Laboratory for Materials Science, Institute of Metal Research, Chinese Academy of Sciences, 110016 Shenyang, China}

\author{Jiantao Wang}
\affiliation{Shenyang National Laboratory for Materials Science, Institute of Metal Research, Chinese Academy of Sciences, 110016 Shenyang, China}

\author{Noah Avargues}
\affiliation{University of Vienna, Faculty of Physics and Center for Computational Materials Science, Kolingasse 14-16, A-1090 Vienna, Austria}

\author{Carla Verdi}
\affiliation{University of Vienna, Faculty of Physics and Center for Computational Materials Science, Kolingasse 14-16, A-1090 Vienna, Austria}

\author{Andreas Singraber}
\affiliation{VASP Software GmbH, Sensengasse 8, A-1090 Vienna, Austria}

\author{Ferenc Karsai}
\affiliation{VASP Software GmbH, Sensengasse 8, A-1090 Vienna, Austria}

\author{Xing-Qiu Chen}
\affiliation{Shenyang National Laboratory for Materials Science, Institute of Metal Research, Chinese Academy of Sciences, 110016 Shenyang, China}

\author{Georg Kresse}
\affiliation{University of Vienna, Faculty of Physics and Center for Computational Materials Science, Kolingasse 14-16, A-1090 Vienna, Austria}
\affiliation{VASP Software GmbH, Sensengasse 8, A-1090 Vienna, Austria}

\begin{abstract}
Adsorption of carbon monoxide (CO) on transition-metal surfaces is a prototypical process
in surface sciences and catalysis. Despite its simplicity, it has posed great challenges to theoretical modeling.
Pretty much all existing density functionals fail to accurately describe surface energies,
CO adsorption site preference, as well as adsorption energies simultaneously.
Although the random phase approximation (RPA) cures these density functional theory failures,
 its large computational cost makes it prohibitive to study the CO adsorption for any but the simplest ordered cases.
 Here, we address these challenges by developing a machine-learned force field (MLFF)
 with near RPA accuracy for the prediction of coverage-dependent  adsorption of CO on the Rh(111) surface
through an efficient on-the-fly active learning procedure and a $\Delta$-machine learning approach.
We show that the RPA-derived MLFF is capable to accurately predict the Rh(111)
surface energy and CO adsorption site preference as well as adsorption energies at different coverages
that are all  in good agreement with experiments.
Moreover, the coverage-dependent ground-state adsorption patterns and adsorption saturation coverage are identified.
\end{abstract}

\maketitle

Carbon monoxide (CO) is one of the important raw materials for organic synthesis
such as the famous Fischer-Tropsch synthesis~\cite{book_Catalysis}.
In these reactions, transition metals usually serve as catalysts and
the catalytic reactions often occur on the transition-metal surfaces.
Therefore, studying CO adsorption on transition-metal surfaces
is technologically and fundamentally of great significance.

However, this prototypical surface science system has posed great challenges to theoretical modeling.
This dates back to the well-known so-called ``CO/Pt(111) puzzle"~\cite{doi:10.1021/jp002302t,doi:10.1063/1.1488596},
which indicates that density functional theory (DFT) using the existing local or semilocal density functionals
tends to favor hollow sites for CO on Pt and Rh, in disagreement with experiments that yield the top site
as the most stable adsorption site~\cite{doi:10.1021/jp002302t,doi:10.1063/1.1488596,
doi:10.1021/acs.jpcc.7b00365,PhysRevB.100.035442,PhysRevB.100.035442,Stroppa_2008,Schimka2010,PhysRevB.83.121410}.
In addition, these density functionals tend to overestimate adsorption energies but underestimate
surface energies~\cite{Stroppa_2008,Schimka2010,PhysRevB.83.121410}.
It has been shown that the failures of local or semilocal density functionals are intimately associated with the incorrect
positioning of the CO frontier orbitals with respect to the Fermi
level~\cite{GIL200371,PhysRevB.83.121410,PhysRevB.69.161401,doi:10.1021/ja0712367,Schimka2010}
and likely caused by density-driven self-interaction errors~\cite{PhysRevB.100.035442}.
As one of the  few parameter-free methods that are able to resolve all these DFT issues,
the random phase approximation (RPA) to the correlation energy~\cite{Ren2012} was shown to
yield predictions not only in qualitative but also in quantitative agreement with experiment~\cite{Schimka2010}.

Presently, most theoretical studies have been restricted to investigating CO adsorption on metal surfaces for well-defined high-symmetry geometries,
while coverage-dependent CO adsorption studies with an exhaustive search of the configuration space are rarely reported.
It is worth mentioning that, using Monte-Carlo simulations based on a classical force field parametrized on DFT,
Shan~\emph{et. al}~\cite{doi:10.1021/jp8094962} pioneeringly studied coverage-dependent CO adsorption on Pt(111).
Nevertheless, the predicted results suffer from the failures of the underlying DFT method.
This leaves several important questions open:
What are the ground-state adsorption patterns at different coverages,
how does the adsorption energy change as a function of coverage,
and what is the saturation coverage?
Here, we set out to address all these issues,
using a combination of machine learning and state-of-the-art methods beyond DFT.
We demonstrate clearly that quantitatively accurate predictions can be obtained by combining these cutting-edge developments into a seamless framework.

In order to accelerate the search of the configuration space,
machine learning based regression techniques provide an elegant solution.
They allow for an efficient and accurate representation of the potential energy surfaces with near
first-principles accuracy~\cite{PhysRevLett.120.026102,PhysRevLett.124.086102,
PhysRevB.99.064114,PhysRevMaterials.2.013803,
doi:10.1021/acs.chemrev.1c00022,doi:10.1021/acs.chemrev.0c00868,doi:10.1021/acs.chemrev.0c01111}.
However, most machine-learned force fields (MLFFs) are based on DFT calculations
and they come with the same limitations as the underlying density functional, as is the case here for CO adsorption on metal surfaces.
In our recent work in Ref.~\cite{PhysRevB.105.L060102},
we proposed a general strategy for generating a MLFF
with beyond-DFT accuracy at a modest computational cost
by combining an efficient on-the-fly active learning method~\cite{PhysRevLett.122.225701,PhysRevB.100.014105}
and a $\Delta$-machine learning ($\Delta$-ML) approach~\cite{PhysRevB.88.054104,Anatole_2015JCTC,Bartokek_SA2017}.
Here, using a similar scheme we develop an RPA-derived MLFF (referred to as MLFF-RPA$^\Delta$)
for the prediction of coverage-dependent CO adsorption on Rh(111). This progresses the state of
art in several key aspects: To machine learn accurate energies for defects or adsorbates is much more challenging than for bulk systems,
since machine learning techniques tend to make relatively larger errors for defects~\cite{PhysRevX.8.041048,PhysRevMaterials.2.013808,PhysRevB.100.144105}.
Furthermore, acquiring the training data is far from trivial,
since the configuration space is huge and in our case not necessarily exhaustively sampled by DFT-based molecular dynamics.
Finally, we need to deal with metallic surfaces and require accurate forces from the employed many-body theory, which puts great demands on the sampling of the Brillouin zone.
There is no {\it a priori} telling whether we can achieve our goal. Nevertheless, we demonstrate convincingly that the MLFF-RPA$^\Delta$ predicts
accurate  Rh(111) surface energy, the correct CO adsorption sites and supremely accurate adsorption energies at all coverages.
All this has not been achieved to date. To put this in context, even relaxations for molecules on surfaces have not been reported for beyond-DFT methods.

We start by introducing the procedure for the construction of the MLFF-RPA$^\Delta$.
Following the scheme proposed in Ref.~\cite{PhysRevB.105.L060102},
the procedure involves building a baseline DFT-derived MLFF
and an intermediate surrogate MLFF model for machine-learning the differences
between the RPA and DFT calculated energies and forces
(called MLFF-$\Delta$ where $\Delta$=RPA$-$DFT).
To this end, we carried out two on-the-fly active learning procedures
based on Bayesian regression~\cite{PhysRevLett.122.225701,PhysRevB.100.014105}
using the separable descriptors~\cite{doi:10.1063/5.0009491}
as implemented in the Vienna \emph{Ab initio} Simulation Package (VASP)~\cite{PhysRevB.47.558, PhysRevB.54.11169}.
First, the baseline DFT-based MLFF was trained on the fly during heating molecular dynamics (MD)
simulations from 0 to 1000 K using typically 20 000 MD steps for many different starting configurations
including bulk Rh and the clean Rh(111) surface as well as different CO adsorption coverages [$\leq$ 0.75 monolayer (ML)].
For the surfaces, six-layer slabs with a vacuum width of 15~$\AA$ and up to $6\times6$ surface unit cells were used.
The DFT calculations were performed using the Perdew-Burke-Ernzerhof (PBE) functional~\cite{PhysRevLett.77.3865}.
Eventually, the training dataset (called $T^A$) contains 5267 structures [see Supplementary Material (SM) Table~S1~\cite{SM}].
Second, another on-the-fly training was performed but using a reduced four-layer slab, $2\times2$ surface unit cells only,
and a smaller vacuum width of 10~$\AA$.
A CUR rank compression ~\cite{Mahoney697,PhysRevB.100.014105}
was applied to this pool of resulting structures, to determine the 100 most representative motifs and structures. These constitute
a second dataset referred to as $T^B$ (SM Table~S2~\cite{SM}). Kernel principal component analysis~\cite{doi:10.1021/acs.accounts.0c00403}
indicates that the sampled motifs cover a wide range of the phase space (SM Fig.~S1~\cite{SM}).

For the structures in $T^B$, the energies and forces were recalculated using the cubic-scaling
finite-temperature RPA~\cite{PhysRevB.90.054115,PhysRevB.94.165109,PhysRevLett.118.106403,PhysRevB.101.205145}.
Partial electronic occupancies and their derivatives are exactly handled, and the first derivatives of the RPA energy
with respect to the positions of the ions can be calculated without approximations at finite electronic temperature.
The MLFF-$\Delta$ was then generated by machine learning the differences in energies and forces
between the RPA and PBE calculations.
We note that for this specific case, learning of the difference may have been somewhat easier
by using a van der Waals (vdW) corrected density functional~\cite{PhysRevLett.92.246401,PhysRevB.81.045401,doi:10.1021/acs.jctc.9b00035}
as the low-level method (SM Table~S5~\cite{SM}); however,
here we desire to prove the effectiveness and generality of the $\Delta$-ML approach by using PBE as the starting point.
Finally, the energies and forces of the structures in $T^A$
were corrected by adding the differences predicted by the MLFF-$\Delta$.
The final MLFF-RPA$^\Delta$ was then fitted to the updated $T^A$.
For more details on the descriptors and training, we refer to SM~\cite{SM}.

\begin{table}
\caption {The validation RMSEs in energies (meV/atom) and forces (eV/\AA)
calculated by KMLFFs and MTPs (see SM~\cite{SM} for details).
}
\begin{ruledtabular}
\begin{tabular}{lcc}
 & Energy & Force \\
 \hline
KMLFF-PBE      &  2.12  &   0.057  \\
MTP-PBE       &   1.71 & 0.044\\
KMLFF-RPA$^\Delta$     &   2.79  &   0.085   \\
MTP-RPA$^\Delta$       & 2.04 & 0.070  \\
\end{tabular}
\end{ruledtabular}
\label{tab:error_MLFF-RPA}
\end{table}

We note that VASP uses a kernel-based method to fit the energies and forces.
Generally, the computational cost for one energy and force calculation scales linearly with respect to
the number of weights and basis functions used in the kernel.
The number of necessary functions scales also unfavorably with respect to the number of species and the size of the configuration space
that one needs to represent accurately.
By contrast, the moment tensor potentials (MTPs) developed by
Shapeev~\cite{Shapeev_MTP2016} are relatively immune to these limitations. They
rely on a linear regression in a space of descriptors that include many-body interactions up to very high order.
MTPs are 1-2 orders of magnitude faster than the kernel-based methods for a comparable accuracy~\cite{doi:10.1021/acs.jpca.9b08723}.
To speed up the calculations, we refitted the on-the-fly generated dataset using MTPs for both, PBE and RPA,
using the MLIP package~\cite{Novikov_2021} (see SM~\cite{SM} for details).
To avoid confusions, the shorthand ``MTP'' refers to the moment tensor potential and ``KMLFF" to the kernel-based approach used in VASP.

\begin{table}
\caption {The Rh(111) surface energy ($E_{\rm s}$) and
CO adsorption energies at different coverages ($E_{\rm ad}$) predicted from first principles and MTPs.
The adopted structure models are given in SM Fig.~S3~\cite{SM}.
}
\begin{ruledtabular}
\begin{tabular}{lccccc}
                           &     PBE  &    MTP-PBE & RPA   &  MTP-RPA$^\Delta$ & Experiment \\
                           \hline
 $E_{\rm s}$ (eV/u.a.)               &    0.813   &     0.809  &      1.027 &      1.020  & 1.039\cite{TYSON1977267} \\
 \hline
 \multicolumn{6}{c}{$E_{\rm ad}$(1/4 ML) (eV/CO)} \\
 Top &   $-$1.916 &  $-$1.885  &   $-$1.464 &   $-$1.482  & 1.45\cite{ABILDPEDERSEN20071747} \\
& & & & &  ($\pm$0.14) \\
fcc &   $-$1.859 &  $-$1.861  &   $-$1.221 &   $-$1.250  &\\
hcp &   $-$1.950 &  $-$1.953  &   $-$1.287 &   $-$1.322  &\\
Bridge &   $-$1.842 &  $-$1.830  &   $-$1.248 &   $-$1.280  &\\
  \hline
 \multicolumn{6}{c}{$E_{\rm ad}$(2/4 ML) (eV/CO)} \\
Top &   $-$1.610 &  $-$1.594  &   $-$1.244 &   $-$1.247  &\\
fcc &   $-$1.618 &  $-$1.607  &   $-$1.085 &   $-$1.097  &\\
hcp &   $-$1.670 &  $-$1.656  &   $-$1.123 &   $-$1.139  &\\
Bridge &   $-$1.739 &  $-$1.724  &   $-$1.269 &   $-$1.271 & \\
  \hline
 \multicolumn{6}{c}{$E_{\rm ad}$(3/4 ML) (eV/CO)} \\
Top &   $-$1.323 &  $-$1.332  &   $-$1.030 &   $-$1.056  &\\
fcc &   $-$1.415 &  $-$1.404  &   $-$0.961 &   $-$0.967  &\\
hcp &   $-$1.446 &  $-$1.425  &   $-$0.985 &   $-$0.982  &\\
Bridge &   $-$1.404 &  $-$1.404  &   $-$1.004 &   $-$0.967  &\\
\end{tabular}
\end{ruledtabular}
\label{tab:Eadsorption}
\end{table}

\begin{figure*}
\begin{center}
\includegraphics[width=1.00\textwidth, clip]{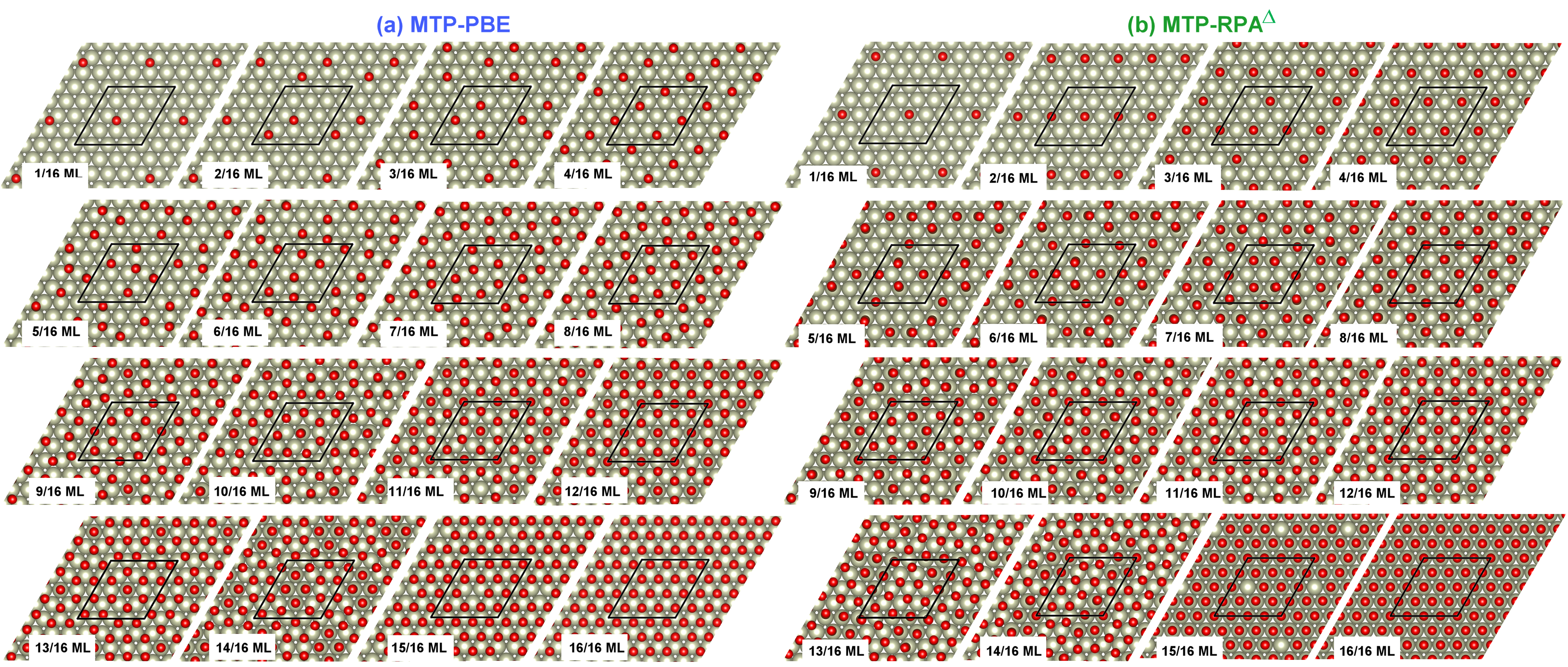}
\end{center}
\caption{Ground-state structures of CO adsorption on Rh(111) for different coverages predicted by (a) MTP-PBE and (b) MTP-RPA$^\Delta$.
The Rh and O atoms are represented as light green and  red balls, respectively. The C atoms (in brown) are underneath the O atoms and thus not visible.
The black lines indicate the $4\times4$ Rh(111) surface that is periodically replicated.
}
\label{fig:MTP_MLPBE_MLRPA_SA}
\end{figure*}

The MLFFs-PBE were validated on a test dataset containing 320 structures
including bulk Rh, the clean Rh(111) surface, and CO adsorption at various coverages (SM Table S3~\cite{SM}).
Because of the large computational cost of RPA calculations,
the validation of MLFFs-RPA$^\Delta$ was restricted to a reduced test dataset including 25 structures of small unit cells (SM Table S4~\cite{SM}).
The root-mean-square errors (RMSEs) are summarized in Table~\ref{tab:error_MLFF-RPA},
and the MLFFs predicted energies and forces against the first-principles data are further illustrated in the SM Fig.~S2~\cite{SM}.
In the present case, the MTPs are slightly more accurate.
The accuracy of the KMLFFs could be improved by increasing the number of basis functions used in the kernel but at the price of increasing the compute time.
Given the better performance of MTPs than the VASP KMLFFs (here a factor of 20-50), we performed the subsequent calculations using MTPs. However, all
main conclusions are unaffected by the choice of the regression method (see SM Table~S5~\cite{SM}).

Next, we test the prediction of physical properties using the MLFFs.
First, the lattice constant of bulk Rh predicted by MTP-PBE is 3.83~$\AA$,
which is in good agreement with the PBE-predicted one (3.82~$\AA$)
and also close to the experimental value (3.80~$\AA$~\cite{PhysRevB.69.075102}). Furthermore,
in line with previous literature results~\cite{Schimka2010}, our RPA-predicted surface energy (1.027 eV/u.a.)
agrees well with the experimental value (1.039 eV/u.a.~\cite{TYSON1977267,VITOS1998186}), whereas
the PBE-predicted one (0.813 eV/u.a.) is  too small (compare Table~\ref{tab:Eadsorption}).
Third, for the CO site preference in the small structures shown in SM Fig.~S3~\cite{SM},
the MTPs predicted adsorption energies are overall in good agreement
with the first-principles data (Table~\ref{tab:Eadsorption}). Note that the relaxed high-symmetry structures
did not enter in the training dataset (except for sampling during the MD), so these are to some extent predictions,
in particular, for the energetically unfavorable sites in DFT. This indicates that the finite-temperature MD sampled
the relevant configuration space sufficiently well.
In addition, the CO site preference for each coverage is also reproduced by the MTPs, except at 3/4 ML.
Taking 1/4 ML for example, both PBE and MTP-PBE yield the site order as hcp $>$ top $>$ fcc $>$ bridge,
whereas RPA and MTP-RPA$^\Delta$ predict the site order: top $>$ hcp $>$ bridge $>$ fcc.
Both RPA- and  MTP-RPA$^\Delta$-predicted site order and adsorption energies are in good agreement
with experiments~\cite{doi:10.1063/1.461508,BEUTLER1997381,WEI199749,doi:10.1063/1.1355767,SMEDH200199,doi:10.1063/1.2184308} (see Table~\ref{tab:Eadsorption}).
By contrast, the PBE- and MTP-PBE-predicted adsorption energies are too large.
This is due to the overestimated binding of CO to the substrate
as evidenced by the  shorter C-Rh bond lengths for PBE (see SM Table~S6~\cite{SM}).
For 3/4 ML, the stability of the bridge site is slightly underestimated by the MLFFs (see Table~\ref{tab:Eadsorption} and SM Table~S5~\cite{SM}),
which we suspect is likely related to insufficient training data. As we will see later, at 3/4 ML a pattern
that maximizes the CO distances mixing hcp, fcc and top sites is preferred (see Fig.~\ref{fig:MTP_MLPBE_MLRPA_SA}).

Having technically validated our MLFFs, we are now in a position to study the coverage-dependent
CO adsorption for less symmetric structures. Using the developed MLFFs, we performed extensive
simulated annealing (SA) MDs for various CO coverages using the LAMMPS code~\cite{LAMMPS}.
We employed a 4$\times$4 slab of six Rh layers with a vacuum width of 15~$\AA$ and  a CO coverage ranging from 1/16 ML to full coverage.
We note that the structures with coverages larger than 0.75 ML were not included in the training dataset.
For each coverage we carried out ten independent SA runs starting from different initial configurations
and for each MD the system was slowly cooled down from 1000 to 0 K using 100 000 MD steps, followed by a structural relaxation.
The configuration with the lowest energy was assumed to be the ground-state structure.

\begin{figure}
\begin{center}
\includegraphics[width=0.40\textwidth,trim = {0.0cm 0.0cm 0.0cm 0.0cm}, clip]{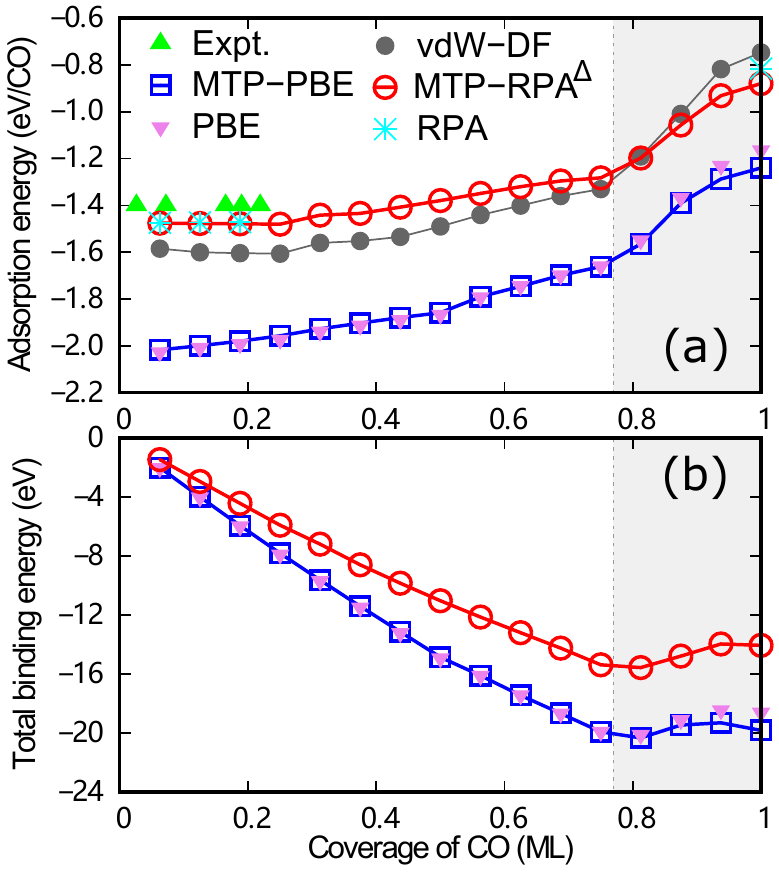}
\end{center}
\caption{(a) CO adsorption energies and (b) total binding energies of the system as a function of CO coverages.
The experimental data are taken from Ref.~\cite{doi:10.1063/1.461508}.
PBE and vdW-DF~\cite{PhysRevLett.92.246401} results
were obtained by relaxing the MTP-PBE and MTP-RPA$^\Delta$ structures, respectively.
Note that the structures with coverages over 0.75 ML (gray shaded region) are not included in the training,
and the RPA calculations were performed using a four-layer slab with a vacuum width of 10~$\AA$.
}
\label{fig:adsorption_vs_ML}
\end{figure}

Figure~\ref{fig:MTP_MLPBE_MLRPA_SA} shows the MTP-predicted coverage-dependent ground-state adsorption patterns
and Fig.~\ref{fig:adsorption_vs_ML} shows the corresponding adsorption energies and
total binding energies as a function of CO coverage.
Let us first focus on the MTP-PBE predictions.
As expected, MTP-PBE has  the same limitations as PBE and tends to
favor the hcp sites for coverages less than or equal to 1/4 ML,
at variance with experiments~\cite{WEI199749,doi:10.1063/1.1355767,SMEDH200199,doi:10.1063/1.2184308}.
The structures are consistent with the principle of hcp site adsorption and avoiding nearest-neighbor adsorption sites.
As the coverage increases to 5/16 ML, CO starts to also adsorb at the fcc sites to maintain a large distance.
Top site adsorption starts to occur from 9/16 ML onward, so maximizing the distance becomes progressively more important
as the coverage increases.
At  0.75 ML, a highly symmetric adsorption pattern
[(2$\times$2)-3CO with equally mixed top, hcp, and fcc sites] is observed. As the coverage continues to increase, the CO
molecules move back to their high-symmetry hcp site, which was also the stable site at lower coverage (also note the vacancy structure
at 15/16 ML).

We verified the predicted adsorption energies by performing PBE relaxations starting from  the optimal structures  (see Fig.~\ref{fig:adsorption_vs_ML}).
Although the agreement is perfect up to 0.75 ML, the errors become larger above 0.75 ML where training data are missing.

Let us now move to the MTP-RPA$^\Delta$ predictions.
MTP-RPA$^\Delta$ restores the correct site order and
predicts that all CO molecules adsorb at the top sites up to 0.5 ML (see Fig.~\ref{fig:MTP_MLPBE_MLRPA_SA}).
The adsorption patterns are clearly different to MTP-PBE at low coverage,
indicating that the lateral interactions are also different. For instance,
CO prefers to adsorb at the third-nearest-neighbor top site, which causes essentially no repulsion (see Fig.~\ref{fig:adsorption_vs_ML}).
At 0.25 ML we now observe a hexagonal instead of a square pattern.
Furthermore, there are some peculiar patterns observed in the structures beyond 0.25 ML:
Three CO molecules have a tendency to form triangles on second-nearest-neighboring Rh atoms.
The predicted tilting toward the center is mostly a result of repulsion from the CO at the nearest-neighbor Rh atoms, as it prevails if a local relaxation using the PBE functional is performed.
The triangular motif  at 5/16 and 6/16 ML is similar to the adsorption pattern for CO on Rh(111) at 1/3 ML~\cite{doi:10.1063/1.1355767}. The experimental $\sqrt{3} \times \sqrt{3}$ structure is found when simulated annealing is performed for e.g. a $3\sqrt{3} \times 3\sqrt{3}$ supercell (see SM Fig.~S4~\cite{SM}).
At 8/16 ML we obtain a square-like structure again with slightly tilted CO molecules. The tilting also prevails during relaxation using PBE.
As the coverage increases further, the threefold hollow sites become occupied, and between 11/16 up to 13/16 ML we observe
similar adsorption patterns as in MTP-PBE. At 0.75 ML, the MTP-RPA$^\Delta$-predicted pattern is
in excellent agreement with the experimental observation~\cite{WEI199749,doi:10.1063/1.1355767,SMEDH200199,doi:10.1063/1.2184308} (SM Fig.~S4~\cite{SM}).
The results at 13/16 and 14/16 ML are also interesting. At 13/16 ML, all CO molecules continue to have six neighbors;
by canting the hexagons relative to the substrate, the CO molecules can move closer together.
At 14/16 ML, most CO molecules still have six neighbors but there are extended defects with pentagonal coordination.
Similar but not identical patterns are also observed for MTP-PBE at coverages larger than 0.75 ML.
At even higher coverage, the CO molecules again move back to the low-energy site (here top),
with vacancy structures occurring before the full coverage is reached.
By performing direct RPA calculations for CO at the four high-symmetry sites at full coverage,
we confirmed that the MTP-RPA$^\Delta$ yields the same site preference as the RPA.

Regarding the coverage-dependent adsorption energies, one can observe that
the adsorption energies decrease as expected with increasing coverage for both MTP-PBE and MTP-RPA$^\Delta$.
However,  the MTP-RPA$^\Delta$-predicted adsorption energies are
generally smaller, in better agreement with the experimental data~\cite{doi:10.1063/1.461508} [see Fig.~\ref{fig:adsorption_vs_ML}(a)].
Also at low coverage, the adsorption energies only marginally change for the MTP-RPA$^\Delta$, in excellent
agreement with experiment~\cite{doi:10.1063/1.461508} and our RPA calculations [see Fig.~\ref{fig:adsorption_vs_ML}(a)]. This is consistent with a very weak CO-CO lateral repulsion. The repulsion is clearly much stronger for the hollow site (PBE curve), where the adsorption energy
already decreases between 0 and 1/4 ML.
By performing local relaxations starting from the MTP-RPA$^\Delta$ predicted structures using PBE,
we have determined that the differences are not related to the functional, but rather related to differences between the substrate mediated
repulsion: For hcp-site adsorption, many neighboring Rh sites are ``poisoned'' resulting in a fairly early decrease of the average adsorption energy.
As the coverage increases further, the CO-CO lateral repulsion increases more steeply.
In particular beyond 14/16 ML, an abrupt decrease of the adsorption energies and binding energies is observed, indicating
that approximately 13/16 ML is the saturation coverage. This is slightly larger than the commonly believed 0.75 ML and achieved by rotating the CO layer with respect to the substrate.
For comparison, we also show the adsorption energies obtained using vdW-DF~\cite{PhysRevLett.92.246401}.
Although vdW-DF predicts the top site to be stable (SM Table S5~\cite{SM}),
it also predicts an increase of the adsorption energy at low coverage, suggesting CO clustering.
As far as we know, clustering has never been observed experimentally and is not confirmed by the RPA either (crosses in Fig.~\ref{fig:adsorption_vs_ML}).

In summary, we have developed an RPA-derived MLFF for predicting the coverage-dependent CO adsorption on Rh(111)
at a modest computational cost through a combination of on-the-fly active learning and $\Delta$-machine learning.
This allows us not only to solve the long-standing CO adsorption puzzle on transition-metal substrates at low coverage,
but also to accurately predict the Rh(111) surface energy, CO adsorption energies, and ground-state adsorption patterns at different coverages.
All these results agree well with experiments, an achievement that is impossible with present density functionals. More broadly speaking,
incorrect structure predictions using approximate density functionals are a common problem. Similar strategies as the one developed here can be used to become truly predictive in chemistry, materials, and surface sciences, including low-dimensional systems, terraces, and step edges.

\begin{acknowledgments}
This work is supported by the Austrian Science Fund (FWF) within the SFB TACO  (Project No. F 81-N)
and the National Natural Science Foundation of China  (No. 52201030).
X.-Q.C. is supported by the National Key R\&D Program of China (No. 2021YFB3501503)
and National Science Fund for Distinguished Young Scholars (No. 51725103).
Supercomputing time on the Vienna Scientific cluster (VSC) is gratefully acknowledged.
\end{acknowledgments}

\bibliographystyle{apsrev4-1}
\bibliography{Reference}

\begin{thebibliography}{67}%
\makeatletter
\providecommand \@ifxundefined [1]{%
 \@ifx{#1\undefined}
}%
\providecommand \@ifnum [1]{%
 \ifnum #1\expandafter \@firstoftwo
 \else \expandafter \@secondoftwo
 \fi
}%
\providecommand \@ifx [1]{%
 \ifx #1\expandafter \@firstoftwo
 \else \expandafter \@secondoftwo
 \fi
}%
\providecommand \natexlab [1]{#1}%
\providecommand \enquote  [1]{``#1''}%
\providecommand \bibnamefont  [1]{#1}%
\providecommand \bibfnamefont [1]{#1}%
\providecommand \citenamefont [1]{#1}%
\providecommand \href@noop [0]{\@secondoftwo}%
\providecommand \href [0]{\begingroup \@sanitize@url \@href}%
\providecommand \@href[1]{\@@startlink{#1}\@@href}%
\providecommand \@@href[1]{\endgroup#1\@@endlink}%
\providecommand \@sanitize@url [0]{\catcode `\\12\catcode `\$12\catcode
  `\&12\catcode `\#12\catcode `\^12\catcode `\_12\catcode `\%12\relax}%
\providecommand \@@startlink[1]{}%
\providecommand \@@endlink[0]{}%
\providecommand \url  [0]{\begingroup\@sanitize@url \@url }%
\providecommand \@url [1]{\endgroup\@href {#1}{\urlprefix }}%
\providecommand \urlprefix  [0]{URL }%
\providecommand \Eprint [0]{\href }%
\providecommand \doibase [0]{http://dx.doi.org/}%
\providecommand \selectlanguage [0]{\@gobble}%
\providecommand \bibinfo  [0]{\@secondoftwo}%
\providecommand \bibfield  [0]{\@secondoftwo}%
\providecommand \translation [1]{[#1]}%
\providecommand \BibitemOpen [0]{}%
\providecommand \bibitemStop [0]{}%
\providecommand \bibitemNoStop [0]{.\EOS\space}%
\providecommand \EOS [0]{\spacefactor3000\relax}%
\providecommand \BibitemShut  [1]{\csname bibitem#1\endcsname}%
\let\auto@bib@innerbib\@empty
\bibitem [{\citenamefont {A.}(1994)}]{book_Catalysis}%
  \BibitemOpen
  \bibfield  {author} {\bibinfo {author} {\bibfnamefont {S.~G.}\ \bibnamefont
  {A.}},\ }\href@noop {} {\emph {\bibinfo {title} {Introduction to Surface
  Chemistry and Catalysis}}}\ (\bibinfo  {publisher} {John Wiley \& Sons Inc,
  New York},\ \bibinfo {year} {1994})\BibitemShut {NoStop}%
\bibitem [{\citenamefont {Feibelman}\ \emph {et~al.}(2001)\citenamefont
  {Feibelman}, \citenamefont {Hammer}, \citenamefont {N{\o}rskov},
  \citenamefont {Wagner}, \citenamefont {Scheffler}, \citenamefont {Stumpf},
  \citenamefont {Watwe},\ and\ \citenamefont
  {Dumesic}}]{doi:10.1021/jp002302t}%
  \BibitemOpen
  \bibfield  {author} {\bibinfo {author} {\bibfnamefont {P.~J.}\ \bibnamefont
  {Feibelman}}, \bibinfo {author} {\bibfnamefont {B.}~\bibnamefont {Hammer}},
  \bibinfo {author} {\bibfnamefont {J.~K.}\ \bibnamefont {N{\o}rskov}},
  \bibinfo {author} {\bibfnamefont {F.}~\bibnamefont {Wagner}}, \bibinfo
  {author} {\bibfnamefont {M.}~\bibnamefont {Scheffler}}, \bibinfo {author}
  {\bibfnamefont {R.}~\bibnamefont {Stumpf}}, \bibinfo {author} {\bibfnamefont
  {R.}~\bibnamefont {Watwe}}, \ and\ \bibinfo {author} {\bibfnamefont
  {J.}~\bibnamefont {Dumesic}},\ }\href {\doibase 10.1021/jp002302t} {\bibfield
   {journal} {\bibinfo  {journal} {The Journal of Physical Chemistry B}\
  }\textbf {\bibinfo {volume} {105}},\ \bibinfo {pages} {4018} (\bibinfo {year}
  {2001})}\BibitemShut {NoStop}%
\bibitem [{\citenamefont {Grinberg}\ \emph {et~al.}(2002)\citenamefont
  {Grinberg}, \citenamefont {Yourdshahyan},\ and\ \citenamefont
  {Rappe}}]{doi:10.1063/1.1488596}%
  \BibitemOpen
  \bibfield  {author} {\bibinfo {author} {\bibfnamefont {I.}~\bibnamefont
  {Grinberg}}, \bibinfo {author} {\bibfnamefont {Y.}~\bibnamefont
  {Yourdshahyan}}, \ and\ \bibinfo {author} {\bibfnamefont {A.~M.}\
  \bibnamefont {Rappe}},\ }\href {\doibase 10.1063/1.1488596} {\bibfield
  {journal} {\bibinfo  {journal} {The Journal of Chemical Physics}\ }\textbf
  {\bibinfo {volume} {117}},\ \bibinfo {pages} {2264} (\bibinfo {year}
  {2002})}\BibitemShut {NoStop}%
\bibitem [{\citenamefont {Janthon}\ \emph {et~al.}(2017)\citenamefont
  {Janthon}, \citenamefont {Vi\~{n}es}, \citenamefont {Sirijaraensre},
  \citenamefont {Limtrakul},\ and\ \citenamefont
  {Illas}}]{doi:10.1021/acs.jpcc.7b00365}%
  \BibitemOpen
  \bibfield  {author} {\bibinfo {author} {\bibfnamefont {P.}~\bibnamefont
  {Janthon}}, \bibinfo {author} {\bibfnamefont {F.}~\bibnamefont {Vi\~{n}es}},
  \bibinfo {author} {\bibfnamefont {J.}~\bibnamefont {Sirijaraensre}}, \bibinfo
  {author} {\bibfnamefont {J.}~\bibnamefont {Limtrakul}}, \ and\ \bibinfo
  {author} {\bibfnamefont {F.}~\bibnamefont {Illas}},\ }\href {\doibase
  10.1021/acs.jpcc.7b00365} {\bibfield  {journal} {\bibinfo  {journal} {The
  Journal of Physical Chemistry C}\ }\textbf {\bibinfo {volume} {121}},\
  \bibinfo {pages} {3970} (\bibinfo {year} {2017})}\BibitemShut {NoStop}%
\bibitem [{\citenamefont {Patra}\ \emph {et~al.}(2019)\citenamefont {Patra},
  \citenamefont {Peng}, \citenamefont {Sun},\ and\ \citenamefont
  {Perdew}}]{PhysRevB.100.035442}%
  \BibitemOpen
  \bibfield  {author} {\bibinfo {author} {\bibfnamefont {A.}~\bibnamefont
  {Patra}}, \bibinfo {author} {\bibfnamefont {H.}~\bibnamefont {Peng}},
  \bibinfo {author} {\bibfnamefont {J.}~\bibnamefont {Sun}}, \ and\ \bibinfo
  {author} {\bibfnamefont {J.~P.}\ \bibnamefont {Perdew}},\ }\href {\doibase
  10.1103/PhysRevB.100.035442} {\bibfield  {journal} {\bibinfo  {journal}
  {Phys. Rev. B}\ }\textbf {\bibinfo {volume} {100}},\ \bibinfo {pages}
  {035442} (\bibinfo {year} {2019})}\BibitemShut {NoStop}%
\bibitem [{\citenamefont {Stroppa}\ and\ \citenamefont
  {Kresse}(2008)}]{Stroppa_2008}%
  \BibitemOpen
  \bibfield  {author} {\bibinfo {author} {\bibfnamefont {A.}~\bibnamefont
  {Stroppa}}\ and\ \bibinfo {author} {\bibfnamefont {G.}~\bibnamefont
  {Kresse}},\ }\href {\doibase 10.1088/1367-2630/10/6/063020} {\bibfield
  {journal} {\bibinfo  {journal} {New Journal of Physics}\ }\textbf {\bibinfo
  {volume} {10}},\ \bibinfo {pages} {063020} (\bibinfo {year}
  {2008})}\BibitemShut {NoStop}%
\bibitem [{\citenamefont {Schimka}\ \emph {et~al.}(2010)\citenamefont
  {Schimka}, \citenamefont {Harl}, \citenamefont {Stroppa}, \citenamefont
  {Gr\"{u}neis}, \citenamefont {Marsman}, \citenamefont {Mittendorfer},\ and\
  \citenamefont {Kresse}}]{Schimka2010}%
  \BibitemOpen
  \bibfield  {author} {\bibinfo {author} {\bibfnamefont {L.}~\bibnamefont
  {Schimka}}, \bibinfo {author} {\bibfnamefont {J.}~\bibnamefont {Harl}},
  \bibinfo {author} {\bibfnamefont {A.}~\bibnamefont {Stroppa}}, \bibinfo
  {author} {\bibfnamefont {A.}~\bibnamefont {Gr\"{u}neis}}, \bibinfo {author}
  {\bibfnamefont {M.}~\bibnamefont {Marsman}}, \bibinfo {author} {\bibfnamefont
  {F.}~\bibnamefont {Mittendorfer}}, \ and\ \bibinfo {author} {\bibfnamefont
  {G.}~\bibnamefont {Kresse}},\ }\href {\doibase 10.1038/nmat2806} {\bibfield
  {journal} {\bibinfo  {journal} {Nature Materials}\ }\textbf {\bibinfo
  {volume} {10}},\ \bibinfo {pages} {741} (\bibinfo {year} {2010})}\BibitemShut
  {NoStop}%
\bibitem [{\citenamefont {Sun}\ \emph {et~al.}(2011)\citenamefont {Sun},
  \citenamefont {Marsman}, \citenamefont {Ruzsinszky}, \citenamefont {Kresse},\
  and\ \citenamefont {Perdew}}]{PhysRevB.83.121410}%
  \BibitemOpen
  \bibfield  {author} {\bibinfo {author} {\bibfnamefont {J.}~\bibnamefont
  {Sun}}, \bibinfo {author} {\bibfnamefont {M.}~\bibnamefont {Marsman}},
  \bibinfo {author} {\bibfnamefont {A.}~\bibnamefont {Ruzsinszky}}, \bibinfo
  {author} {\bibfnamefont {G.}~\bibnamefont {Kresse}}, \ and\ \bibinfo {author}
  {\bibfnamefont {J.~P.}\ \bibnamefont {Perdew}},\ }\href {\doibase
  10.1103/PhysRevB.83.121410} {\bibfield  {journal} {\bibinfo  {journal} {Phys.
  Rev. B}\ }\textbf {\bibinfo {volume} {83}},\ \bibinfo {pages} {121410}
  (\bibinfo {year} {2011})}\BibitemShut {NoStop}%
\bibitem [{\citenamefont {Gil}\ \emph {et~al.}(2003)\citenamefont {Gil},
  \citenamefont {Clotet}, \citenamefont {Ricart}, \citenamefont {Kresse},
  \citenamefont {Garc\'{\i}a-Hern\'{a}ndez}, \citenamefont {R\"{o}sch},\ and\
  \citenamefont {Sautet}}]{GIL200371}%
  \BibitemOpen
  \bibfield  {author} {\bibinfo {author} {\bibfnamefont {A.}~\bibnamefont
  {Gil}}, \bibinfo {author} {\bibfnamefont {A.}~\bibnamefont {Clotet}},
  \bibinfo {author} {\bibfnamefont {J.~M.}\ \bibnamefont {Ricart}}, \bibinfo
  {author} {\bibfnamefont {G.}~\bibnamefont {Kresse}}, \bibinfo {author}
  {\bibfnamefont {M.}~\bibnamefont {Garc\'{\i}a-Hern\'{a}ndez}}, \bibinfo
  {author} {\bibfnamefont {N.}~\bibnamefont {R\"{o}sch}}, \ and\ \bibinfo
  {author} {\bibfnamefont {P.}~\bibnamefont {Sautet}},\ }\href {\doibase
  https://doi.org/10.1016/S0039-6028(03)00307-8} {\bibfield  {journal}
  {\bibinfo  {journal} {Surface Science}\ }\textbf {\bibinfo {volume} {530}},\
  \bibinfo {pages} {71} (\bibinfo {year} {2003})}\BibitemShut {NoStop}%
\bibitem [{\citenamefont {Mason}\ \emph {et~al.}(2004)\citenamefont {Mason},
  \citenamefont {Grinberg},\ and\ \citenamefont {Rappe}}]{PhysRevB.69.161401}%
  \BibitemOpen
  \bibfield  {author} {\bibinfo {author} {\bibfnamefont {S.~E.}\ \bibnamefont
  {Mason}}, \bibinfo {author} {\bibfnamefont {I.}~\bibnamefont {Grinberg}}, \
  and\ \bibinfo {author} {\bibfnamefont {A.~M.}\ \bibnamefont {Rappe}},\ }\href
  {\doibase 10.1103/PhysRevB.69.161401} {\bibfield  {journal} {\bibinfo
  {journal} {Phys. Rev. B}\ }\textbf {\bibinfo {volume} {69}},\ \bibinfo
  {pages} {161401} (\bibinfo {year} {2004})}\BibitemShut {NoStop}%
\bibitem [{\citenamefont {Wang}\ \emph {et~al.}(2007)\citenamefont {Wang},
  \citenamefont {de~Gironcoli}, \citenamefont {Hush},\ and\ \citenamefont
  {Reimers}}]{doi:10.1021/ja0712367}%
  \BibitemOpen
  \bibfield  {author} {\bibinfo {author} {\bibfnamefont {Y.}~\bibnamefont
  {Wang}}, \bibinfo {author} {\bibfnamefont {S.}~\bibnamefont {de~Gironcoli}},
  \bibinfo {author} {\bibfnamefont {N.~S.}\ \bibnamefont {Hush}}, \ and\
  \bibinfo {author} {\bibfnamefont {J.~R.}\ \bibnamefont {Reimers}},\ }\href
  {\doibase 10.1021/ja0712367} {\bibfield  {journal} {\bibinfo  {journal}
  {Journal of the American Chemical Society}\ }\textbf {\bibinfo {volume}
  {129}},\ \bibinfo {pages} {10402} (\bibinfo {year} {2007})}\BibitemShut
  {NoStop}%
\bibitem [{\citenamefont {Ren}\ \emph {et~al.}(2012)\citenamefont {Ren},
  \citenamefont {Rinke}, \citenamefont {Joas},\ and\ \citenamefont
  {Scheffler}}]{Ren2012}%
  \BibitemOpen
  \bibfield  {author} {\bibinfo {author} {\bibfnamefont {X.}~\bibnamefont
  {Ren}}, \bibinfo {author} {\bibfnamefont {P.}~\bibnamefont {Rinke}}, \bibinfo
  {author} {\bibfnamefont {C.}~\bibnamefont {Joas}}, \ and\ \bibinfo {author}
  {\bibfnamefont {M.}~\bibnamefont {Scheffler}},\ }\href {\doibase
  10.1007/s10853-012-6570-4} {\bibfield  {journal} {\bibinfo  {journal}
  {Journal of Materials Science}\ }\textbf {\bibinfo {volume} {47}},\ \bibinfo
  {pages} {7447} (\bibinfo {year} {2012})}\BibitemShut {NoStop}%
\bibitem [{\citenamefont {Shan}\ \emph {et~al.}(2009)\citenamefont {Shan},
  \citenamefont {Zhao}, \citenamefont {Hyun}, \citenamefont {Kapur},
  \citenamefont {Nicholas},\ and\ \citenamefont {Cho}}]{doi:10.1021/jp8094962}%
  \BibitemOpen
  \bibfield  {author} {\bibinfo {author} {\bibfnamefont {B.}~\bibnamefont
  {Shan}}, \bibinfo {author} {\bibfnamefont {Y.}~\bibnamefont {Zhao}}, \bibinfo
  {author} {\bibfnamefont {J.}~\bibnamefont {Hyun}}, \bibinfo {author}
  {\bibfnamefont {N.}~\bibnamefont {Kapur}}, \bibinfo {author} {\bibfnamefont
  {J.~B.}\ \bibnamefont {Nicholas}}, \ and\ \bibinfo {author} {\bibfnamefont
  {K.}~\bibnamefont {Cho}},\ }\href {\doibase 10.1021/jp8094962} {\bibfield
  {journal} {\bibinfo  {journal} {The Journal of Physical Chemistry C}\
  }\textbf {\bibinfo {volume} {113}},\ \bibinfo {pages} {6088} (\bibinfo {year}
  {2009})}\BibitemShut {NoStop}%
\bibitem [{\citenamefont {Jacobsen}\ \emph {et~al.}(2018)\citenamefont
  {Jacobsen}, \citenamefont {J\o{}rgensen},\ and\ \citenamefont
  {Hammer}}]{PhysRevLett.120.026102}%
  \BibitemOpen
  \bibfield  {author} {\bibinfo {author} {\bibfnamefont {T.~L.}\ \bibnamefont
  {Jacobsen}}, \bibinfo {author} {\bibfnamefont {M.~S.}\ \bibnamefont
  {J\o{}rgensen}}, \ and\ \bibinfo {author} {\bibfnamefont {B.}~\bibnamefont
  {Hammer}},\ }\href {\doibase 10.1103/PhysRevLett.120.026102} {\bibfield
  {journal} {\bibinfo  {journal} {Phys. Rev. Lett.}\ }\textbf {\bibinfo
  {volume} {120}},\ \bibinfo {pages} {026102} (\bibinfo {year}
  {2018})}\BibitemShut {NoStop}%
\bibitem [{\citenamefont {Bisbo}\ and\ \citenamefont
  {Hammer}(2020)}]{PhysRevLett.124.086102}%
  \BibitemOpen
  \bibfield  {author} {\bibinfo {author} {\bibfnamefont {M.~K.}\ \bibnamefont
  {Bisbo}}\ and\ \bibinfo {author} {\bibfnamefont {B.}~\bibnamefont {Hammer}},\
  }\href {\doibase 10.1103/PhysRevLett.124.086102} {\bibfield  {journal}
  {\bibinfo  {journal} {Phys. Rev. Lett.}\ }\textbf {\bibinfo {volume} {124}},\
  \bibinfo {pages} {086102} (\bibinfo {year} {2020})}\BibitemShut {NoStop}%
\bibitem [{\citenamefont {Podryabinkin}\ \emph {et~al.}(2019)\citenamefont
  {Podryabinkin}, \citenamefont {Tikhonov}, \citenamefont {Shapeev},\ and\
  \citenamefont {Oganov}}]{PhysRevB.99.064114}%
  \BibitemOpen
  \bibfield  {author} {\bibinfo {author} {\bibfnamefont {E.~V.}\ \bibnamefont
  {Podryabinkin}}, \bibinfo {author} {\bibfnamefont {E.~V.}\ \bibnamefont
  {Tikhonov}}, \bibinfo {author} {\bibfnamefont {A.~V.}\ \bibnamefont
  {Shapeev}}, \ and\ \bibinfo {author} {\bibfnamefont {A.~R.}\ \bibnamefont
  {Oganov}},\ }\href {\doibase 10.1103/PhysRevB.99.064114} {\bibfield
  {journal} {\bibinfo  {journal} {Phys. Rev. B}\ }\textbf {\bibinfo {volume}
  {99}},\ \bibinfo {pages} {064114} (\bibinfo {year} {2019})}\BibitemShut
  {NoStop}%
\bibitem [{\citenamefont {Yamashita}\ \emph {et~al.}(2018)\citenamefont
  {Yamashita}, \citenamefont {Sato}, \citenamefont {Kino}, \citenamefont
  {Miyake}, \citenamefont {Tsuda},\ and\ \citenamefont
  {Oguchi}}]{PhysRevMaterials.2.013803}%
  \BibitemOpen
  \bibfield  {author} {\bibinfo {author} {\bibfnamefont {T.}~\bibnamefont
  {Yamashita}}, \bibinfo {author} {\bibfnamefont {N.}~\bibnamefont {Sato}},
  \bibinfo {author} {\bibfnamefont {H.}~\bibnamefont {Kino}}, \bibinfo {author}
  {\bibfnamefont {T.}~\bibnamefont {Miyake}}, \bibinfo {author} {\bibfnamefont
  {K.}~\bibnamefont {Tsuda}}, \ and\ \bibinfo {author} {\bibfnamefont
  {T.}~\bibnamefont {Oguchi}},\ }\href {\doibase
  10.1103/PhysRevMaterials.2.013803} {\bibfield  {journal} {\bibinfo  {journal}
  {Phys. Rev. Materials}\ }\textbf {\bibinfo {volume} {2}},\ \bibinfo {pages}
  {013803} (\bibinfo {year} {2018})}\BibitemShut {NoStop}%
\bibitem [{\citenamefont {Deringer}\ \emph {et~al.}(2021)\citenamefont
  {Deringer}, \citenamefont {Bart\'{o}k}, \citenamefont {Bernstein},
  \citenamefont {Wilkins}, \citenamefont {Ceriotti},\ and\ \citenamefont
  {Cs\'{a}nyi}}]{doi:10.1021/acs.chemrev.1c00022}%
  \BibitemOpen
  \bibfield  {author} {\bibinfo {author} {\bibfnamefont {V.~L.}\ \bibnamefont
  {Deringer}}, \bibinfo {author} {\bibfnamefont {A.~P.}\ \bibnamefont
  {Bart\'{o}k}}, \bibinfo {author} {\bibfnamefont {N.}~\bibnamefont
  {Bernstein}}, \bibinfo {author} {\bibfnamefont {D.~M.}\ \bibnamefont
  {Wilkins}}, \bibinfo {author} {\bibfnamefont {M.}~\bibnamefont {Ceriotti}}, \
  and\ \bibinfo {author} {\bibfnamefont {G.}~\bibnamefont {Cs\'{a}nyi}},\
  }\href {\doibase 10.1021/acs.chemrev.1c00022} {\bibfield  {journal} {\bibinfo
   {journal} {Chemical Reviews}\ }\textbf {\bibinfo {volume} {121}},\ \bibinfo
  {pages} {10073} (\bibinfo {year} {2021})}\BibitemShut {NoStop}%
\bibitem [{\citenamefont {Behler}(2021)}]{doi:10.1021/acs.chemrev.0c00868}%
  \BibitemOpen
  \bibfield  {author} {\bibinfo {author} {\bibfnamefont {J.}~\bibnamefont
  {Behler}},\ }\href {\doibase 10.1021/acs.chemrev.0c00868} {\bibfield
  {journal} {\bibinfo  {journal} {Chemical Reviews}\ }\textbf {\bibinfo
  {volume} {121}},\ \bibinfo {pages} {10037} (\bibinfo {year}
  {2021})}\BibitemShut {NoStop}%
\bibitem [{\citenamefont {Unke}\ \emph {et~al.}(2021)\citenamefont {Unke},
  \citenamefont {Chmiela}, \citenamefont {Sauceda}, \citenamefont {Gastegger},
  \citenamefont {Poltavsky}, \citenamefont {Sch\"{u}tt}, \citenamefont
  {Tkatchenko},\ and\ \citenamefont
  {M\"{u}ller}}]{doi:10.1021/acs.chemrev.0c01111}%
  \BibitemOpen
  \bibfield  {author} {\bibinfo {author} {\bibfnamefont {O.~T.}\ \bibnamefont
  {Unke}}, \bibinfo {author} {\bibfnamefont {S.}~\bibnamefont {Chmiela}},
  \bibinfo {author} {\bibfnamefont {H.~E.}\ \bibnamefont {Sauceda}}, \bibinfo
  {author} {\bibfnamefont {M.}~\bibnamefont {Gastegger}}, \bibinfo {author}
  {\bibfnamefont {I.}~\bibnamefont {Poltavsky}}, \bibinfo {author}
  {\bibfnamefont {K.~T.}\ \bibnamefont {Sch\"{u}tt}}, \bibinfo {author}
  {\bibfnamefont {A.}~\bibnamefont {Tkatchenko}}, \ and\ \bibinfo {author}
  {\bibfnamefont {K.-R.}\ \bibnamefont {M\"{u}ller}},\ }\href {\doibase
  10.1021/acs.chemrev.0c01111} {\bibfield  {journal} {\bibinfo  {journal}
  {Chemical Reviews}\ }\textbf {\bibinfo {volume} {121}},\ \bibinfo {pages}
  {10142} (\bibinfo {year} {2021})}\BibitemShut {NoStop}%
\bibitem [{\citenamefont {Liu}\ \emph {et~al.}(2022)\citenamefont {Liu},
  \citenamefont {Verdi}, \citenamefont {Karsai},\ and\ \citenamefont
  {Kresse}}]{PhysRevB.105.L060102}%
  \BibitemOpen
  \bibfield  {author} {\bibinfo {author} {\bibfnamefont {P.}~\bibnamefont
  {Liu}}, \bibinfo {author} {\bibfnamefont {C.}~\bibnamefont {Verdi}}, \bibinfo
  {author} {\bibfnamefont {F.}~\bibnamefont {Karsai}}, \ and\ \bibinfo {author}
  {\bibfnamefont {G.}~\bibnamefont {Kresse}},\ }\href {\doibase
  10.1103/PhysRevB.105.L060102} {\bibfield  {journal} {\bibinfo  {journal}
  {Phys. Rev. B}\ }\textbf {\bibinfo {volume} {105}},\ \bibinfo {pages}
  {L060102} (\bibinfo {year} {2022})}\BibitemShut {NoStop}%
\bibitem [{\citenamefont {Jinnouchi}\ \emph
  {et~al.}(2019{\natexlab{a}})\citenamefont {Jinnouchi}, \citenamefont
  {Lahnsteiner}, \citenamefont {Karsai}, \citenamefont {Kresse},\ and\
  \citenamefont {Bokdam}}]{PhysRevLett.122.225701}%
  \BibitemOpen
  \bibfield  {author} {\bibinfo {author} {\bibfnamefont {R.}~\bibnamefont
  {Jinnouchi}}, \bibinfo {author} {\bibfnamefont {J.}~\bibnamefont
  {Lahnsteiner}}, \bibinfo {author} {\bibfnamefont {F.}~\bibnamefont {Karsai}},
  \bibinfo {author} {\bibfnamefont {G.}~\bibnamefont {Kresse}}, \ and\ \bibinfo
  {author} {\bibfnamefont {M.}~\bibnamefont {Bokdam}},\ }\href {\doibase
  10.1103/PhysRevLett.122.225701} {\bibfield  {journal} {\bibinfo  {journal}
  {Phys. Rev. Lett.}\ }\textbf {\bibinfo {volume} {122}},\ \bibinfo {pages}
  {225701} (\bibinfo {year} {2019}{\natexlab{a}})}\BibitemShut {NoStop}%
\bibitem [{\citenamefont {Jinnouchi}\ \emph
  {et~al.}(2019{\natexlab{b}})\citenamefont {Jinnouchi}, \citenamefont
  {Karsai},\ and\ \citenamefont {Kresse}}]{PhysRevB.100.014105}%
  \BibitemOpen
  \bibfield  {author} {\bibinfo {author} {\bibfnamefont {R.}~\bibnamefont
  {Jinnouchi}}, \bibinfo {author} {\bibfnamefont {F.}~\bibnamefont {Karsai}}, \
  and\ \bibinfo {author} {\bibfnamefont {G.}~\bibnamefont {Kresse}},\ }\href
  {\doibase 10.1103/PhysRevB.100.014105} {\bibfield  {journal} {\bibinfo
  {journal} {Phys. Rev. B}\ }\textbf {\bibinfo {volume} {100}},\ \bibinfo
  {pages} {014105} (\bibinfo {year} {2019}{\natexlab{b}})}\BibitemShut
  {NoStop}%
\bibitem [{\citenamefont {Bart\'ok}\ \emph
  {et~al.}(2013{\natexlab{a}})\citenamefont {Bart\'ok}, \citenamefont {Gillan},
  \citenamefont {Manby},\ and\ \citenamefont {Cs\'anyi}}]{PhysRevB.88.054104}%
  \BibitemOpen
  \bibfield  {author} {\bibinfo {author} {\bibfnamefont {A.~P.}\ \bibnamefont
  {Bart\'ok}}, \bibinfo {author} {\bibfnamefont {M.~J.}\ \bibnamefont
  {Gillan}}, \bibinfo {author} {\bibfnamefont {F.~R.}\ \bibnamefont {Manby}}, \
  and\ \bibinfo {author} {\bibfnamefont {G.}~\bibnamefont {Cs\'anyi}},\ }\href
  {\doibase 10.1103/PhysRevB.88.054104} {\bibfield  {journal} {\bibinfo
  {journal} {Phys. Rev. B}\ }\textbf {\bibinfo {volume} {88}},\ \bibinfo
  {pages} {054104} (\bibinfo {year} {2013}{\natexlab{a}})}\BibitemShut
  {NoStop}%
\bibitem [{\citenamefont {Ramakrishnan}\ \emph {et~al.}(2015)\citenamefont
  {Ramakrishnan}, \citenamefont {Dral}, \citenamefont {Rupp},\ and\
  \citenamefont {von Lilienfeld}}]{Anatole_2015JCTC}%
  \BibitemOpen
  \bibfield  {author} {\bibinfo {author} {\bibfnamefont {R.}~\bibnamefont
  {Ramakrishnan}}, \bibinfo {author} {\bibfnamefont {P.~O.}\ \bibnamefont
  {Dral}}, \bibinfo {author} {\bibfnamefont {M.}~\bibnamefont {Rupp}}, \ and\
  \bibinfo {author} {\bibfnamefont {O.~A.}\ \bibnamefont {von Lilienfeld}},\
  }\href {\doibase 10.1021/acs.jctc.5b00099} {\bibfield  {journal} {\bibinfo
  {journal} {J. Chem. Theory Comput.}\ }\textbf {\bibinfo {volume} {11}},\
  \bibinfo {pages} {2087} (\bibinfo {year} {2015})}\BibitemShut {NoStop}%
\bibitem [{\citenamefont {Bart{\'o}k}\ \emph {et~al.}(2017)\citenamefont
  {Bart{\'o}k}, \citenamefont {De}, \citenamefont {Poelking}, \citenamefont
  {Bernstein}, \citenamefont {Kermode}, \citenamefont {Cs{\'a}nyi},\ and\
  \citenamefont {Ceriotti}}]{Bartokek_SA2017}%
  \BibitemOpen
  \bibfield  {author} {\bibinfo {author} {\bibfnamefont {A.~P.}\ \bibnamefont
  {Bart{\'o}k}}, \bibinfo {author} {\bibfnamefont {S.}~\bibnamefont {De}},
  \bibinfo {author} {\bibfnamefont {C.}~\bibnamefont {Poelking}}, \bibinfo
  {author} {\bibfnamefont {N.}~\bibnamefont {Bernstein}}, \bibinfo {author}
  {\bibfnamefont {J.~R.}\ \bibnamefont {Kermode}}, \bibinfo {author}
  {\bibfnamefont {G.}~\bibnamefont {Cs{\'a}nyi}}, \ and\ \bibinfo {author}
  {\bibfnamefont {M.}~\bibnamefont {Ceriotti}},\ }\href
  {https://advances.sciencemag.org/content/3/12/e1701816} {\bibfield  {journal}
  {\bibinfo  {journal} {Science Advances}\ }\textbf {\bibinfo {volume} {3}},\
  \bibinfo {pages} {e1701816} (\bibinfo {year} {2017})}\BibitemShut {NoStop}%
\bibitem [{\citenamefont {Bart\'ok}\ \emph {et~al.}(2018)\citenamefont
  {Bart\'ok}, \citenamefont {Kermode}, \citenamefont {Bernstein},\ and\
  \citenamefont {Cs\'anyi}}]{PhysRevX.8.041048}%
  \BibitemOpen
  \bibfield  {author} {\bibinfo {author} {\bibfnamefont {A.~P.}\ \bibnamefont
  {Bart\'ok}}, \bibinfo {author} {\bibfnamefont {J.}~\bibnamefont {Kermode}},
  \bibinfo {author} {\bibfnamefont {N.}~\bibnamefont {Bernstein}}, \ and\
  \bibinfo {author} {\bibfnamefont {G.}~\bibnamefont {Cs\'anyi}},\ }\href
  {\doibase 10.1103/PhysRevX.8.041048} {\bibfield  {journal} {\bibinfo
  {journal} {Phys. Rev. X}\ }\textbf {\bibinfo {volume} {8}},\ \bibinfo {pages}
  {041048} (\bibinfo {year} {2018})}\BibitemShut {NoStop}%
\bibitem [{\citenamefont {Dragoni}\ \emph {et~al.}(2018)\citenamefont
  {Dragoni}, \citenamefont {Daff}, \citenamefont {Cs\'anyi},\ and\
  \citenamefont {Marzari}}]{PhysRevMaterials.2.013808}%
  \BibitemOpen
  \bibfield  {author} {\bibinfo {author} {\bibfnamefont {D.}~\bibnamefont
  {Dragoni}}, \bibinfo {author} {\bibfnamefont {T.~D.}\ \bibnamefont {Daff}},
  \bibinfo {author} {\bibfnamefont {G.}~\bibnamefont {Cs\'anyi}}, \ and\
  \bibinfo {author} {\bibfnamefont {N.}~\bibnamefont {Marzari}},\ }\href
  {\doibase 10.1103/PhysRevMaterials.2.013808} {\bibfield  {journal} {\bibinfo
  {journal} {Phys. Rev. Materials}\ }\textbf {\bibinfo {volume} {2}},\ \bibinfo
  {pages} {013808} (\bibinfo {year} {2018})}\BibitemShut {NoStop}%
\bibitem [{\citenamefont {Byggm\"astar}\ \emph {et~al.}(2019)\citenamefont
  {Byggm\"astar}, \citenamefont {Hamedani}, \citenamefont {Nordlund},\ and\
  \citenamefont {Djurabekova}}]{PhysRevB.100.144105}%
  \BibitemOpen
  \bibfield  {author} {\bibinfo {author} {\bibfnamefont {J.}~\bibnamefont
  {Byggm\"astar}}, \bibinfo {author} {\bibfnamefont {A.}~\bibnamefont
  {Hamedani}}, \bibinfo {author} {\bibfnamefont {K.}~\bibnamefont {Nordlund}},
  \ and\ \bibinfo {author} {\bibfnamefont {F.}~\bibnamefont {Djurabekova}},\
  }\href {\doibase 10.1103/PhysRevB.100.144105} {\bibfield  {journal} {\bibinfo
   {journal} {Phys. Rev. B}\ }\textbf {\bibinfo {volume} {100}},\ \bibinfo
  {pages} {144105} (\bibinfo {year} {2019})}\BibitemShut {NoStop}%
\bibitem [{\citenamefont {Jinnouchi}\ \emph
  {et~al.}(2020{\natexlab{a}})\citenamefont {Jinnouchi}, \citenamefont
  {Karsai}, \citenamefont {Verdi}, \citenamefont {Asahi},\ and\ \citenamefont
  {Kresse}}]{doi:10.1063/5.0009491}%
  \BibitemOpen
  \bibfield  {author} {\bibinfo {author} {\bibfnamefont {R.}~\bibnamefont
  {Jinnouchi}}, \bibinfo {author} {\bibfnamefont {F.}~\bibnamefont {Karsai}},
  \bibinfo {author} {\bibfnamefont {C.}~\bibnamefont {Verdi}}, \bibinfo
  {author} {\bibfnamefont {R.}~\bibnamefont {Asahi}}, \ and\ \bibinfo {author}
  {\bibfnamefont {G.}~\bibnamefont {Kresse}},\ }\href {\doibase
  10.1063/5.0009491} {\bibfield  {journal} {\bibinfo  {journal} {J. Chem.
  Phys.}\ }\textbf {\bibinfo {volume} {152}},\ \bibinfo {pages} {234102}
  (\bibinfo {year} {2020}{\natexlab{a}})}\BibitemShut {NoStop}%
\bibitem [{\citenamefont {Kresse}\ and\ \citenamefont
  {Hafner}(1993)}]{PhysRevB.47.558}%
  \BibitemOpen
  \bibfield  {author} {\bibinfo {author} {\bibfnamefont {G.}~\bibnamefont
  {Kresse}}\ and\ \bibinfo {author} {\bibfnamefont {J.}~\bibnamefont
  {Hafner}},\ }\href {\doibase 10.1103/PhysRevB.47.558} {\bibfield  {journal}
  {\bibinfo  {journal} {Phys. Rev. B}\ }\textbf {\bibinfo {volume} {47}},\
  \bibinfo {pages} {558} (\bibinfo {year} {1993})}\BibitemShut {NoStop}%
\bibitem [{\citenamefont {Kresse}\ and\ \citenamefont
  {Furthm\"uller}(1996)}]{PhysRevB.54.11169}%
  \BibitemOpen
  \bibfield  {author} {\bibinfo {author} {\bibfnamefont {G.}~\bibnamefont
  {Kresse}}\ and\ \bibinfo {author} {\bibfnamefont {J.}~\bibnamefont
  {Furthm\"uller}},\ }\href {\doibase 10.1103/PhysRevB.54.11169} {\bibfield
  {journal} {\bibinfo  {journal} {Phys. Rev. B}\ }\textbf {\bibinfo {volume}
  {54}},\ \bibinfo {pages} {11169} (\bibinfo {year} {1996})}\BibitemShut
  {NoStop}%
\bibitem [{\citenamefont {Perdew}\ \emph {et~al.}(1996)\citenamefont {Perdew},
  \citenamefont {Burke},\ and\ \citenamefont
  {Ernzerhof}}]{PhysRevLett.77.3865}%
  \BibitemOpen
  \bibfield  {author} {\bibinfo {author} {\bibfnamefont {J.~P.}\ \bibnamefont
  {Perdew}}, \bibinfo {author} {\bibfnamefont {K.}~\bibnamefont {Burke}}, \
  and\ \bibinfo {author} {\bibfnamefont {M.}~\bibnamefont {Ernzerhof}},\ }\href
  {\doibase 10.1103/PhysRevLett.77.3865} {\bibfield  {journal} {\bibinfo
  {journal} {Phys. Rev. Lett.}\ }\textbf {\bibinfo {volume} {77}},\ \bibinfo
  {pages} {3865} (\bibinfo {year} {1996})}\BibitemShut {NoStop}%
\bibitem [{SM()}]{SM}%
  \BibitemOpen
  \href@noop {} {\emph {\bibinfo {title} {\rm See Supplemental Material for the
  details of first-principles calculations, MLFF training and validation, and
  kernel principal component analysis of local atomic environments as well as
  the structural models used for calculating the CO adsorption energies in
  Table~\ref{tab:Eadsorption}, which include
  Refs.~\cite{PhysRevB.50.17953,PhysRevB.59.1758,Peitao_2020,Carla_2021,
  PhysRevB.101.060201,book_Allen_Tildesley,PhysRevLett.45.1196,PhysRevB.87.184115,doi:10.1063/1.3382344,PhysRevB.59.7413}}}}\BibitemShut
  {NoStop}%
\bibitem [{\citenamefont {Mahoney}\ and\ \citenamefont
  {Drineas}(2009)}]{Mahoney697}%
  \BibitemOpen
  \bibfield  {author} {\bibinfo {author} {\bibfnamefont {M.~W.}\ \bibnamefont
  {Mahoney}}\ and\ \bibinfo {author} {\bibfnamefont {P.}~\bibnamefont
  {Drineas}},\ }\href {\doibase 10.1073/pnas.0803205106} {\bibfield  {journal}
  {\bibinfo  {journal} {Proc. Natl. Acad. Sci. U.S.A.}\ }\textbf {\bibinfo
  {volume} {106}},\ \bibinfo {pages} {697} (\bibinfo {year}
  {2009})}\BibitemShut {NoStop}%
\bibitem [{\citenamefont {Cheng}\ \emph {et~al.}(2020)\citenamefont {Cheng},
  \citenamefont {Griffiths}, \citenamefont {Wengert}, \citenamefont {Kunkel},
  \citenamefont {Stenczel}, \citenamefont {Zhu}, \citenamefont {Deringer},
  \citenamefont {Bernstein}, \citenamefont {Margraf}, \citenamefont {Reuter},\
  and\ \citenamefont {Csanyi}}]{doi:10.1021/acs.accounts.0c00403}%
  \BibitemOpen
  \bibfield  {author} {\bibinfo {author} {\bibfnamefont {B.}~\bibnamefont
  {Cheng}}, \bibinfo {author} {\bibfnamefont {R.-R.}\ \bibnamefont
  {Griffiths}}, \bibinfo {author} {\bibfnamefont {S.}~\bibnamefont {Wengert}},
  \bibinfo {author} {\bibfnamefont {C.}~\bibnamefont {Kunkel}}, \bibinfo
  {author} {\bibfnamefont {T.}~\bibnamefont {Stenczel}}, \bibinfo {author}
  {\bibfnamefont {B.}~\bibnamefont {Zhu}}, \bibinfo {author} {\bibfnamefont
  {V.~L.}\ \bibnamefont {Deringer}}, \bibinfo {author} {\bibfnamefont
  {N.}~\bibnamefont {Bernstein}}, \bibinfo {author} {\bibfnamefont {J.~T.}\
  \bibnamefont {Margraf}}, \bibinfo {author} {\bibfnamefont {K.}~\bibnamefont
  {Reuter}}, \ and\ \bibinfo {author} {\bibfnamefont {G.}~\bibnamefont
  {Csanyi}},\ }\href {\doibase 10.1021/acs.accounts.0c00403} {\bibfield
  {journal} {\bibinfo  {journal} {Accounts of Chemical Research}\ }\textbf
  {\bibinfo {volume} {53}},\ \bibinfo {pages} {1981} (\bibinfo {year}
  {2020})}\BibitemShut {NoStop}%
\bibitem [{\citenamefont {Kaltak}\ \emph {et~al.}(2014)\citenamefont {Kaltak},
  \citenamefont {Klime\v{s}},\ and\ \citenamefont
  {Kresse}}]{PhysRevB.90.054115}%
  \BibitemOpen
  \bibfield  {author} {\bibinfo {author} {\bibfnamefont {M.}~\bibnamefont
  {Kaltak}}, \bibinfo {author} {\bibfnamefont {J.}~\bibnamefont {Klime\v{s}}},
  \ and\ \bibinfo {author} {\bibfnamefont {G.}~\bibnamefont {Kresse}},\ }\href
  {\doibase 10.1103/PhysRevB.90.054115} {\bibfield  {journal} {\bibinfo
  {journal} {Phys. Rev. B}\ }\textbf {\bibinfo {volume} {90}},\ \bibinfo
  {pages} {054115} (\bibinfo {year} {2014})}\BibitemShut {NoStop}%
\bibitem [{\citenamefont {Liu}\ \emph {et~al.}(2016)\citenamefont {Liu},
  \citenamefont {Kaltak}, \citenamefont {Klime\v{s}},\ and\ \citenamefont
  {Kresse}}]{PhysRevB.94.165109}%
  \BibitemOpen
  \bibfield  {author} {\bibinfo {author} {\bibfnamefont {P.}~\bibnamefont
  {Liu}}, \bibinfo {author} {\bibfnamefont {M.}~\bibnamefont {Kaltak}},
  \bibinfo {author} {\bibfnamefont {J.}~\bibnamefont {Klime\v{s}}}, \ and\
  \bibinfo {author} {\bibfnamefont {G.}~\bibnamefont {Kresse}},\ }\href
  {\doibase 10.1103/PhysRevB.94.165109} {\bibfield  {journal} {\bibinfo
  {journal} {Phys. Rev. B}\ }\textbf {\bibinfo {volume} {94}},\ \bibinfo
  {pages} {165109} (\bibinfo {year} {2016})}\BibitemShut {NoStop}%
\bibitem [{\citenamefont {Ramberger}\ \emph {et~al.}(2017)\citenamefont
  {Ramberger}, \citenamefont {Sch\"afer},\ and\ \citenamefont
  {Kresse}}]{PhysRevLett.118.106403}%
  \BibitemOpen
  \bibfield  {author} {\bibinfo {author} {\bibfnamefont {B.}~\bibnamefont
  {Ramberger}}, \bibinfo {author} {\bibfnamefont {T.}~\bibnamefont
  {Sch\"afer}}, \ and\ \bibinfo {author} {\bibfnamefont {G.}~\bibnamefont
  {Kresse}},\ }\href {\doibase 10.1103/PhysRevLett.118.106403} {\bibfield
  {journal} {\bibinfo  {journal} {Phys. Rev. Lett.}\ }\textbf {\bibinfo
  {volume} {118}},\ \bibinfo {pages} {106403} (\bibinfo {year}
  {2017})}\BibitemShut {NoStop}%
\bibitem [{\citenamefont {Kaltak}\ and\ \citenamefont
  {Kresse}(2020)}]{PhysRevB.101.205145}%
  \BibitemOpen
  \bibfield  {author} {\bibinfo {author} {\bibfnamefont {M.}~\bibnamefont
  {Kaltak}}\ and\ \bibinfo {author} {\bibfnamefont {G.}~\bibnamefont
  {Kresse}},\ }\href {\doibase 10.1103/PhysRevB.101.205145} {\bibfield
  {journal} {\bibinfo  {journal} {Phys. Rev. B}\ }\textbf {\bibinfo {volume}
  {101}},\ \bibinfo {pages} {205145} (\bibinfo {year} {2020})}\BibitemShut
  {NoStop}%
\bibitem [{\citenamefont {Dion}\ \emph {et~al.}(2004)\citenamefont {Dion},
  \citenamefont {Rydberg}, \citenamefont {Schr\"oder}, \citenamefont
  {Langreth},\ and\ \citenamefont {Lundqvist}}]{PhysRevLett.92.246401}%
  \BibitemOpen
  \bibfield  {author} {\bibinfo {author} {\bibfnamefont {M.}~\bibnamefont
  {Dion}}, \bibinfo {author} {\bibfnamefont {H.}~\bibnamefont {Rydberg}},
  \bibinfo {author} {\bibfnamefont {E.}~\bibnamefont {Schr\"oder}}, \bibinfo
  {author} {\bibfnamefont {D.~C.}\ \bibnamefont {Langreth}}, \ and\ \bibinfo
  {author} {\bibfnamefont {B.~I.}\ \bibnamefont {Lundqvist}},\ }\href {\doibase
  10.1103/PhysRevLett.92.246401} {\bibfield  {journal} {\bibinfo  {journal}
  {Phys. Rev. Lett.}\ }\textbf {\bibinfo {volume} {92}},\ \bibinfo {pages}
  {246401} (\bibinfo {year} {2004})}\BibitemShut {NoStop}%
\bibitem [{\citenamefont {Lazi\ifmmode~\acute{c}\else \'{c}\fi{}}\ \emph
  {et~al.}(2010)\citenamefont {Lazi\ifmmode~\acute{c}\else \'{c}\fi{}},
  \citenamefont {Alaei}, \citenamefont {Atodiresei}, \citenamefont {Caciuc},
  \citenamefont {Brako},\ and\ \citenamefont {Bl\"ugel}}]{PhysRevB.81.045401}%
  \BibitemOpen
  \bibfield  {author} {\bibinfo {author} {\bibfnamefont {P.}~\bibnamefont
  {Lazi\ifmmode~\acute{c}\else \'{c}\fi{}}}, \bibinfo {author} {\bibfnamefont
  {M.}~\bibnamefont {Alaei}}, \bibinfo {author} {\bibfnamefont
  {N.}~\bibnamefont {Atodiresei}}, \bibinfo {author} {\bibfnamefont
  {V.}~\bibnamefont {Caciuc}}, \bibinfo {author} {\bibfnamefont
  {R.}~\bibnamefont {Brako}}, \ and\ \bibinfo {author} {\bibfnamefont
  {S.}~\bibnamefont {Bl\"ugel}},\ }\href {\doibase 10.1103/PhysRevB.81.045401}
  {\bibfield  {journal} {\bibinfo  {journal} {Phys. Rev. B}\ }\textbf {\bibinfo
  {volume} {81}},\ \bibinfo {pages} {045401} (\bibinfo {year}
  {2010})}\BibitemShut {NoStop}%
\bibitem [{\citenamefont {Mahlberg}\ \emph {et~al.}(2019)\citenamefont
  {Mahlberg}, \citenamefont {Sakong}, \citenamefont {Forster-Tonigold},\ and\
  \citenamefont {Groß}}]{doi:10.1021/acs.jctc.9b00035}%
  \BibitemOpen
  \bibfield  {author} {\bibinfo {author} {\bibfnamefont {D.}~\bibnamefont
  {Mahlberg}}, \bibinfo {author} {\bibfnamefont {S.}~\bibnamefont {Sakong}},
  \bibinfo {author} {\bibfnamefont {K.}~\bibnamefont {Forster-Tonigold}}, \
  and\ \bibinfo {author} {\bibfnamefont {A.}~\bibnamefont {Groß}},\ }\href
  {\doibase 10.1021/acs.jctc.9b00035} {\bibfield  {journal} {\bibinfo
  {journal} {Journal of Chemical Theory and Computation}\ }\textbf {\bibinfo
  {volume} {15}},\ \bibinfo {pages} {3250} (\bibinfo {year}
  {2019})}\BibitemShut {NoStop}%
\bibitem [{\citenamefont {Shapeev}(2016)}]{Shapeev_MTP2016}%
  \BibitemOpen
  \bibfield  {author} {\bibinfo {author} {\bibfnamefont {A.~V.}\ \bibnamefont
  {Shapeev}},\ }\href {\doibase 10.1137/15M1054183} {\bibfield  {journal}
  {\bibinfo  {journal} {Multiscale Modeling \& Simulation}\ }\textbf {\bibinfo
  {volume} {14}},\ \bibinfo {pages} {1153} (\bibinfo {year}
  {2016})}\BibitemShut {NoStop}%
\bibitem [{\citenamefont {Zuo}\ \emph {et~al.}(2020)\citenamefont {Zuo},
  \citenamefont {Chen}, \citenamefont {Li}, \citenamefont {Deng}, \citenamefont
  {Chen}, \citenamefont {Behler}, \citenamefont {Csányi}, \citenamefont
  {Shapeev}, \citenamefont {Thompson}, \citenamefont {Wood},\ and\
  \citenamefont {Ong}}]{doi:10.1021/acs.jpca.9b08723}%
  \BibitemOpen
  \bibfield  {author} {\bibinfo {author} {\bibfnamefont {Y.}~\bibnamefont
  {Zuo}}, \bibinfo {author} {\bibfnamefont {C.}~\bibnamefont {Chen}}, \bibinfo
  {author} {\bibfnamefont {X.}~\bibnamefont {Li}}, \bibinfo {author}
  {\bibfnamefont {Z.}~\bibnamefont {Deng}}, \bibinfo {author} {\bibfnamefont
  {Y.}~\bibnamefont {Chen}}, \bibinfo {author} {\bibfnamefont {J.}~\bibnamefont
  {Behler}}, \bibinfo {author} {\bibfnamefont {G.}~\bibnamefont {Csányi}},
  \bibinfo {author} {\bibfnamefont {A.~V.}\ \bibnamefont {Shapeev}}, \bibinfo
  {author} {\bibfnamefont {A.~P.}\ \bibnamefont {Thompson}}, \bibinfo {author}
  {\bibfnamefont {M.~A.}\ \bibnamefont {Wood}}, \ and\ \bibinfo {author}
  {\bibfnamefont {S.~P.}\ \bibnamefont {Ong}},\ }\href {\doibase
  10.1021/acs.jpca.9b08723} {\bibfield  {journal} {\bibinfo  {journal} {The
  Journal of Physical Chemistry A}\ }\textbf {\bibinfo {volume} {124}},\
  \bibinfo {pages} {731} (\bibinfo {year} {2020})}\BibitemShut {NoStop}%
\bibitem [{\citenamefont {Novikov}\ \emph {et~al.}(2021)\citenamefont
  {Novikov}, \citenamefont {Gubaev}, \citenamefont {Podryabinkin},\ and\
  \citenamefont {Shapeev}}]{Novikov_2021}%
  \BibitemOpen
  \bibfield  {author} {\bibinfo {author} {\bibfnamefont {I.~S.}\ \bibnamefont
  {Novikov}}, \bibinfo {author} {\bibfnamefont {K.}~\bibnamefont {Gubaev}},
  \bibinfo {author} {\bibfnamefont {E.~V.}\ \bibnamefont {Podryabinkin}}, \
  and\ \bibinfo {author} {\bibfnamefont {A.~V.}\ \bibnamefont {Shapeev}},\
  }\href {\doibase 10.1088/2632-2153/abc9fe} {\bibfield  {journal} {\bibinfo
  {journal} {Machine Learning: Science and Technology}\ }\textbf {\bibinfo
  {volume} {2}},\ \bibinfo {pages} {025002} (\bibinfo {year}
  {2021})}\BibitemShut {NoStop}%
\bibitem [{\citenamefont {Tyson}\ and\ \citenamefont
  {Miller}(1977)}]{TYSON1977267}%
  \BibitemOpen
  \bibfield  {author} {\bibinfo {author} {\bibfnamefont {W.}~\bibnamefont
  {Tyson}}\ and\ \bibinfo {author} {\bibfnamefont {W.}~\bibnamefont {Miller}},\
  }\href {\doibase https://doi.org/10.1016/0039-6028(77)90442-3} {\bibfield
  {journal} {\bibinfo  {journal} {Surface Science}\ }\textbf {\bibinfo {volume}
  {62}},\ \bibinfo {pages} {267} (\bibinfo {year} {1977})}\BibitemShut
  {NoStop}%
\bibitem [{\citenamefont {Abild-Pedersen}\ and\ \citenamefont
  {Andersson}(2007)}]{ABILDPEDERSEN20071747}%
  \BibitemOpen
  \bibfield  {author} {\bibinfo {author} {\bibfnamefont {F.}~\bibnamefont
  {Abild-Pedersen}}\ and\ \bibinfo {author} {\bibfnamefont {M.}~\bibnamefont
  {Andersson}},\ }\href {\doibase https://doi.org/10.1016/j.susc.2007.01.052}
  {\bibfield  {journal} {\bibinfo  {journal} {Surface Science}\ }\textbf
  {\bibinfo {volume} {601}},\ \bibinfo {pages} {1747} (\bibinfo {year}
  {2007})}\BibitemShut {NoStop}%
\bibitem [{\citenamefont {Staroverov}\ \emph {et~al.}(2004)\citenamefont
  {Staroverov}, \citenamefont {Scuseria}, \citenamefont {Tao},\ and\
  \citenamefont {Perdew}}]{PhysRevB.69.075102}%
  \BibitemOpen
  \bibfield  {author} {\bibinfo {author} {\bibfnamefont {V.~N.}\ \bibnamefont
  {Staroverov}}, \bibinfo {author} {\bibfnamefont {G.~E.}\ \bibnamefont
  {Scuseria}}, \bibinfo {author} {\bibfnamefont {J.}~\bibnamefont {Tao}}, \
  and\ \bibinfo {author} {\bibfnamefont {J.~P.}\ \bibnamefont {Perdew}},\
  }\href {\doibase 10.1103/PhysRevB.69.075102} {\bibfield  {journal} {\bibinfo
  {journal} {Phys. Rev. B}\ }\textbf {\bibinfo {volume} {69}},\ \bibinfo
  {pages} {075102} (\bibinfo {year} {2004})}\BibitemShut {NoStop}%
\bibitem [{\citenamefont {Vitos}\ \emph {et~al.}(1998)\citenamefont {Vitos},
  \citenamefont {Ruban}, \citenamefont {Skriver},\ and\ \citenamefont
  {Koll\'{a}r}}]{VITOS1998186}%
  \BibitemOpen
  \bibfield  {author} {\bibinfo {author} {\bibfnamefont {L.}~\bibnamefont
  {Vitos}}, \bibinfo {author} {\bibfnamefont {A.}~\bibnamefont {Ruban}},
  \bibinfo {author} {\bibfnamefont {H.}~\bibnamefont {Skriver}}, \ and\
  \bibinfo {author} {\bibfnamefont {J.}~\bibnamefont {Koll\'{a}r}},\ }\href
  {\doibase https://doi.org/10.1016/S0039-6028(98)00363-X} {\bibfield
  {journal} {\bibinfo  {journal} {Surface Science}\ }\textbf {\bibinfo {volume}
  {411}},\ \bibinfo {pages} {186} (\bibinfo {year} {1998})}\BibitemShut
  {NoStop}%
\bibitem [{\citenamefont {Peterlinz}\ \emph {et~al.}(1991)\citenamefont
  {Peterlinz}, \citenamefont {Curtiss},\ and\ \citenamefont
  {Sibener}}]{doi:10.1063/1.461508}%
  \BibitemOpen
  \bibfield  {author} {\bibinfo {author} {\bibfnamefont {K.~A.}\ \bibnamefont
  {Peterlinz}}, \bibinfo {author} {\bibfnamefont {T.~J.}\ \bibnamefont
  {Curtiss}}, \ and\ \bibinfo {author} {\bibfnamefont {S.~J.}\ \bibnamefont
  {Sibener}},\ }\href {\doibase 10.1063/1.461508} {\bibfield  {journal}
  {\bibinfo  {journal} {The Journal of Chemical Physics}\ }\textbf {\bibinfo
  {volume} {95}},\ \bibinfo {pages} {6972} (\bibinfo {year}
  {1991})}\BibitemShut {NoStop}%
\bibitem [{\citenamefont {Beutler}\ \emph {et~al.}(1997)\citenamefont
  {Beutler}, \citenamefont {Lundgren}, \citenamefont {Nyholm}, \citenamefont
  {Andersen}, \citenamefont {Setlik},\ and\ \citenamefont
  {Heskett}}]{BEUTLER1997381}%
  \BibitemOpen
  \bibfield  {author} {\bibinfo {author} {\bibfnamefont {A.}~\bibnamefont
  {Beutler}}, \bibinfo {author} {\bibfnamefont {E.}~\bibnamefont {Lundgren}},
  \bibinfo {author} {\bibfnamefont {R.}~\bibnamefont {Nyholm}}, \bibinfo
  {author} {\bibfnamefont {J.}~\bibnamefont {Andersen}}, \bibinfo {author}
  {\bibfnamefont {B.}~\bibnamefont {Setlik}}, \ and\ \bibinfo {author}
  {\bibfnamefont {D.}~\bibnamefont {Heskett}},\ }\href {\doibase
  https://doi.org/10.1016/S0039-6028(96)01014-X} {\bibfield  {journal}
  {\bibinfo  {journal} {Surface Science}\ }\textbf {\bibinfo {volume} {371}},\
  \bibinfo {pages} {381} (\bibinfo {year} {1997})}\BibitemShut {NoStop}%
\bibitem [{\citenamefont {Wei}\ \emph {et~al.}(1997)\citenamefont {Wei},
  \citenamefont {Skelton},\ and\ \citenamefont {Kevan}}]{WEI199749}%
  \BibitemOpen
  \bibfield  {author} {\bibinfo {author} {\bibfnamefont {D.}~\bibnamefont
  {Wei}}, \bibinfo {author} {\bibfnamefont {D.}~\bibnamefont {Skelton}}, \ and\
  \bibinfo {author} {\bibfnamefont {S.}~\bibnamefont {Kevan}},\ }\href
  {\doibase https://doi.org/10.1016/S0039-6028(97)00091-5} {\bibfield
  {journal} {\bibinfo  {journal} {Surface Science}\ }\textbf {\bibinfo {volume}
  {381}},\ \bibinfo {pages} {49} (\bibinfo {year} {1997})}\BibitemShut
  {NoStop}%
\bibitem [{\citenamefont {Linke}\ \emph {et~al.}(2001)\citenamefont {Linke},
  \citenamefont {Curulla}, \citenamefont {Hopstaken},\ and\ \citenamefont
  {Niemantsverdriet}}]{doi:10.1063/1.1355767}%
  \BibitemOpen
  \bibfield  {author} {\bibinfo {author} {\bibfnamefont {R.}~\bibnamefont
  {Linke}}, \bibinfo {author} {\bibfnamefont {D.}~\bibnamefont {Curulla}},
  \bibinfo {author} {\bibfnamefont {M.~J.~P.}\ \bibnamefont {Hopstaken}}, \
  and\ \bibinfo {author} {\bibfnamefont {J.~W.}\ \bibnamefont
  {Niemantsverdriet}},\ }\href {\doibase 10.1063/1.1355767} {\bibfield
  {journal} {\bibinfo  {journal} {The Journal of Chemical Physics}\ }\textbf
  {\bibinfo {volume} {115}},\ \bibinfo {pages} {8209} (\bibinfo {year}
  {2001})}\BibitemShut {NoStop}%
\bibitem [{\citenamefont {Smedh}\ \emph {et~al.}(2001)\citenamefont {Smedh},
  \citenamefont {Beutler}, \citenamefont {Ramsvik}, \citenamefont {Nyholm},
  \citenamefont {Borg}, \citenamefont {Andersen}, \citenamefont {Duschek},
  \citenamefont {Sock}, \citenamefont {Netzer},\ and\ \citenamefont
  {Ramsey}}]{SMEDH200199}%
  \BibitemOpen
  \bibfield  {author} {\bibinfo {author} {\bibfnamefont {M.}~\bibnamefont
  {Smedh}}, \bibinfo {author} {\bibfnamefont {A.}~\bibnamefont {Beutler}},
  \bibinfo {author} {\bibfnamefont {T.}~\bibnamefont {Ramsvik}}, \bibinfo
  {author} {\bibfnamefont {R.}~\bibnamefont {Nyholm}}, \bibinfo {author}
  {\bibfnamefont {M.}~\bibnamefont {Borg}}, \bibinfo {author} {\bibfnamefont
  {J.}~\bibnamefont {Andersen}}, \bibinfo {author} {\bibfnamefont
  {R.}~\bibnamefont {Duschek}}, \bibinfo {author} {\bibfnamefont
  {M.}~\bibnamefont {Sock}}, \bibinfo {author} {\bibfnamefont {F.}~\bibnamefont
  {Netzer}}, \ and\ \bibinfo {author} {\bibfnamefont {M.}~\bibnamefont
  {Ramsey}},\ }\href {\doibase https://doi.org/10.1016/S0039-6028(01)01357-7}
  {\bibfield  {journal} {\bibinfo  {journal} {Surface Science}\ }\textbf
  {\bibinfo {volume} {491}},\ \bibinfo {pages} {99} (\bibinfo {year}
  {2001})}\BibitemShut {NoStop}%
\bibitem [{\citenamefont {Krenn}\ \emph {et~al.}(2006)\citenamefont {Krenn},
  \citenamefont {Bako},\ and\ \citenamefont
  {Schennach}}]{doi:10.1063/1.2184308}%
  \BibitemOpen
  \bibfield  {author} {\bibinfo {author} {\bibfnamefont {G.}~\bibnamefont
  {Krenn}}, \bibinfo {author} {\bibfnamefont {I.}~\bibnamefont {Bako}}, \ and\
  \bibinfo {author} {\bibfnamefont {R.}~\bibnamefont {Schennach}},\ }\href
  {\doibase 10.1063/1.2184308} {\bibfield  {journal} {\bibinfo  {journal} {The
  Journal of Chemical Physics}\ }\textbf {\bibinfo {volume} {124}},\ \bibinfo
  {pages} {144703} (\bibinfo {year} {2006})}\BibitemShut {NoStop}%
\bibitem [{\citenamefont {Thompson}\ \emph {et~al.}(2022)\citenamefont
  {Thompson}, \citenamefont {Aktulga}, \citenamefont {Berger}, \citenamefont
  {Bolintineanu}, \citenamefont {Brown}, \citenamefont {Crozier}, \citenamefont
  {in~'t Veld}, \citenamefont {Kohlmeyer}, \citenamefont {Moore}, \citenamefont
  {Nguyen}, \citenamefont {Shan}, \citenamefont {Stevens}, \citenamefont
  {Tranchida}, \citenamefont {Trott},\ and\ \citenamefont {Plimpton}}]{LAMMPS}%
  \BibitemOpen
  \bibfield  {author} {\bibinfo {author} {\bibfnamefont {A.~P.}\ \bibnamefont
  {Thompson}}, \bibinfo {author} {\bibfnamefont {H.~M.}\ \bibnamefont
  {Aktulga}}, \bibinfo {author} {\bibfnamefont {R.}~\bibnamefont {Berger}},
  \bibinfo {author} {\bibfnamefont {D.~S.}\ \bibnamefont {Bolintineanu}},
  \bibinfo {author} {\bibfnamefont {W.~M.}\ \bibnamefont {Brown}}, \bibinfo
  {author} {\bibfnamefont {P.~S.}\ \bibnamefont {Crozier}}, \bibinfo {author}
  {\bibfnamefont {P.~J.}\ \bibnamefont {in~'t Veld}}, \bibinfo {author}
  {\bibfnamefont {A.}~\bibnamefont {Kohlmeyer}}, \bibinfo {author}
  {\bibfnamefont {S.~G.}\ \bibnamefont {Moore}}, \bibinfo {author}
  {\bibfnamefont {T.~D.}\ \bibnamefont {Nguyen}}, \bibinfo {author}
  {\bibfnamefont {R.}~\bibnamefont {Shan}}, \bibinfo {author} {\bibfnamefont
  {M.~J.}\ \bibnamefont {Stevens}}, \bibinfo {author} {\bibfnamefont
  {J.}~\bibnamefont {Tranchida}}, \bibinfo {author} {\bibfnamefont
  {C.}~\bibnamefont {Trott}}, \ and\ \bibinfo {author} {\bibfnamefont {S.~J.}\
  \bibnamefont {Plimpton}},\ }\href {\doibase 10.1016/j.cpc.2021.108171}
  {\bibfield  {journal} {\bibinfo  {journal} {Comp. Phys. Comm.}\ }\textbf
  {\bibinfo {volume} {271}},\ \bibinfo {pages} {108171} (\bibinfo {year}
  {2022})}\BibitemShut {NoStop}%
\bibitem [{\citenamefont {Bl\"ochl}(1994)}]{PhysRevB.50.17953}%
  \BibitemOpen
  \bibfield  {author} {\bibinfo {author} {\bibfnamefont {P.~E.}\ \bibnamefont
  {Bl\"ochl}},\ }\href {\doibase 10.1103/PhysRevB.50.17953} {\bibfield
  {journal} {\bibinfo  {journal} {Phys. Rev. B}\ }\textbf {\bibinfo {volume}
  {50}},\ \bibinfo {pages} {17953} (\bibinfo {year} {1994})}\BibitemShut
  {NoStop}%
\bibitem [{\citenamefont {Kresse}\ and\ \citenamefont
  {Joubert}(1999)}]{PhysRevB.59.1758}%
  \BibitemOpen
  \bibfield  {author} {\bibinfo {author} {\bibfnamefont {G.}~\bibnamefont
  {Kresse}}\ and\ \bibinfo {author} {\bibfnamefont {D.}~\bibnamefont
  {Joubert}},\ }\href {\doibase 10.1103/PhysRevB.59.1758} {\bibfield  {journal}
  {\bibinfo  {journal} {Phys. Rev. B}\ }\textbf {\bibinfo {volume} {59}},\
  \bibinfo {pages} {1758} (\bibinfo {year} {1999})}\BibitemShut {NoStop}%
\bibitem [{\citenamefont {Liu}\ \emph {et~al.}(2021)\citenamefont {Liu},
  \citenamefont {Verdi}, \citenamefont {Karsai},\ and\ \citenamefont
  {Kresse}}]{Peitao_2020}%
  \BibitemOpen
  \bibfield  {author} {\bibinfo {author} {\bibfnamefont {P.}~\bibnamefont
  {Liu}}, \bibinfo {author} {\bibfnamefont {C.}~\bibnamefont {Verdi}}, \bibinfo
  {author} {\bibfnamefont {F.}~\bibnamefont {Karsai}}, \ and\ \bibinfo {author}
  {\bibfnamefont {G.}~\bibnamefont {Kresse}},\ }\href {\doibase
  10.1103/PhysRevMaterials.5.053804} {\bibfield  {journal} {\bibinfo  {journal}
  {Phys. Rev. Materials}\ }\textbf {\bibinfo {volume} {5}},\ \bibinfo {pages}
  {053804} (\bibinfo {year} {2021})}\BibitemShut {NoStop}%
\bibitem [{\citenamefont {Verdi}\ \emph {et~al.}(2021)\citenamefont {Verdi},
  \citenamefont {Karsai}, \citenamefont {Liu}, \citenamefont {Jinnouchi},\ and\
  \citenamefont {Kresse}}]{Carla_2021}%
  \BibitemOpen
  \bibfield  {author} {\bibinfo {author} {\bibfnamefont {C.}~\bibnamefont
  {Verdi}}, \bibinfo {author} {\bibfnamefont {F.}~\bibnamefont {Karsai}},
  \bibinfo {author} {\bibfnamefont {P.}~\bibnamefont {Liu}}, \bibinfo {author}
  {\bibfnamefont {R.}~\bibnamefont {Jinnouchi}}, \ and\ \bibinfo {author}
  {\bibfnamefont {G.}~\bibnamefont {Kresse}},\ }\href {\doibase
  10.1038/s41524-021-00630-5} {\bibfield  {journal} {\bibinfo  {journal} {npj
  Computational Materials}\ }\textbf {\bibinfo {volume} {7}},\ \bibinfo {pages}
  {156} (\bibinfo {year} {2021})}\BibitemShut {NoStop}%
\bibitem [{\citenamefont {Jinnouchi}\ \emph
  {et~al.}(2020{\natexlab{b}})\citenamefont {Jinnouchi}, \citenamefont
  {Karsai},\ and\ \citenamefont {Kresse}}]{PhysRevB.101.060201}%
  \BibitemOpen
  \bibfield  {author} {\bibinfo {author} {\bibfnamefont {R.}~\bibnamefont
  {Jinnouchi}}, \bibinfo {author} {\bibfnamefont {F.}~\bibnamefont {Karsai}}, \
  and\ \bibinfo {author} {\bibfnamefont {G.}~\bibnamefont {Kresse}},\ }\href
  {\doibase 10.1103/PhysRevB.101.060201} {\bibfield  {journal} {\bibinfo
  {journal} {Phys. Rev. B}\ }\textbf {\bibinfo {volume} {101}},\ \bibinfo
  {pages} {060201} (\bibinfo {year} {2020}{\natexlab{b}})}\BibitemShut
  {NoStop}%
\bibitem [{\citenamefont {Allen}\ and\ \citenamefont
  {Tildesley}(1991)}]{book_Allen_Tildesley}%
  \BibitemOpen
  \bibfield  {author} {\bibinfo {author} {\bibfnamefont {M.~P.}\ \bibnamefont
  {Allen}}\ and\ \bibinfo {author} {\bibfnamefont {D.~J.}\ \bibnamefont
  {Tildesley}},\ }\href@noop {} {\emph {\bibinfo {title} {Computer simulation
  of liquids (Oxford university press: New York)}}}\ (\bibinfo {year}
  {1991})\BibitemShut {NoStop}%
\bibitem [{\citenamefont {Parrinello}\ and\ \citenamefont
  {Rahman}(1980)}]{PhysRevLett.45.1196}%
  \BibitemOpen
  \bibfield  {author} {\bibinfo {author} {\bibfnamefont {M.}~\bibnamefont
  {Parrinello}}\ and\ \bibinfo {author} {\bibfnamefont {A.}~\bibnamefont
  {Rahman}},\ }\href {\doibase 10.1103/PhysRevLett.45.1196} {\bibfield
  {journal} {\bibinfo  {journal} {Phys. Rev. Lett.}\ }\textbf {\bibinfo
  {volume} {45}},\ \bibinfo {pages} {1196} (\bibinfo {year}
  {1980})}\BibitemShut {NoStop}%
\bibitem [{\citenamefont {Bart\'ok}\ \emph
  {et~al.}(2013{\natexlab{b}})\citenamefont {Bart\'ok}, \citenamefont
  {Kondor},\ and\ \citenamefont {Cs\'anyi}}]{PhysRevB.87.184115}%
  \BibitemOpen
  \bibfield  {author} {\bibinfo {author} {\bibfnamefont {A.~P.}\ \bibnamefont
  {Bart\'ok}}, \bibinfo {author} {\bibfnamefont {R.}~\bibnamefont {Kondor}}, \
  and\ \bibinfo {author} {\bibfnamefont {G.}~\bibnamefont {Cs\'anyi}},\ }\href
  {\doibase 10.1103/PhysRevB.87.184115} {\bibfield  {journal} {\bibinfo
  {journal} {Phys. Rev. B}\ }\textbf {\bibinfo {volume} {87}},\ \bibinfo
  {pages} {184115} (\bibinfo {year} {2013}{\natexlab{b}})}\BibitemShut
  {NoStop}%
\bibitem [{\citenamefont {Grimme}\ \emph {et~al.}(2010)\citenamefont {Grimme},
  \citenamefont {Antony}, \citenamefont {Ehrlich},\ and\ \citenamefont
  {Krieg}}]{doi:10.1063/1.3382344}%
  \BibitemOpen
  \bibfield  {author} {\bibinfo {author} {\bibfnamefont {S.}~\bibnamefont
  {Grimme}}, \bibinfo {author} {\bibfnamefont {J.}~\bibnamefont {Antony}},
  \bibinfo {author} {\bibfnamefont {S.}~\bibnamefont {Ehrlich}}, \ and\
  \bibinfo {author} {\bibfnamefont {H.}~\bibnamefont {Krieg}},\ }\href
  {\doibase 10.1063/1.3382344} {\bibfield  {journal} {\bibinfo  {journal} {The
  Journal of Chemical Physics}\ }\textbf {\bibinfo {volume} {132}},\ \bibinfo
  {pages} {154104} (\bibinfo {year} {2010})}\BibitemShut {NoStop}%
\bibitem [{\citenamefont {Hammer}\ \emph {et~al.}(1999)\citenamefont {Hammer},
  \citenamefont {Hansen},\ and\ \citenamefont {N\o{}rskov}}]{PhysRevB.59.7413}%
  \BibitemOpen
  \bibfield  {author} {\bibinfo {author} {\bibfnamefont {B.}~\bibnamefont
  {Hammer}}, \bibinfo {author} {\bibfnamefont {L.~B.}\ \bibnamefont {Hansen}},
  \ and\ \bibinfo {author} {\bibfnamefont {J.~K.}\ \bibnamefont {N\o{}rskov}},\
  }\href {\doibase 10.1103/PhysRevB.59.7413} {\bibfield  {journal} {\bibinfo
  {journal} {Phys. Rev. B}\ }\textbf {\bibinfo {volume} {59}},\ \bibinfo
  {pages} {7413} (\bibinfo {year} {1999})}\BibitemShut {NoStop}%
\end{thebibliography}%


\begin{thebibliography}{30}%
\makeatletter
\providecommand \@ifxundefined [1]{%
 \@ifx{#1\undefined}
}%
\providecommand \@ifnum [1]{%
 \ifnum #1\expandafter \@firstoftwo
 \else \expandafter \@secondoftwo
 \fi
}%
\providecommand \@ifx [1]{%
 \ifx #1\expandafter \@firstoftwo
 \else \expandafter \@secondoftwo
 \fi
}%
\providecommand \natexlab [1]{#1}%
\providecommand \enquote  [1]{``#1''}%
\providecommand \bibnamefont  [1]{#1}%
\providecommand \bibfnamefont [1]{#1}%
\providecommand \citenamefont [1]{#1}%
\providecommand \href@noop [0]{\@secondoftwo}%
\providecommand \href [0]{\begingroup \@sanitize@url \@href}%
\providecommand \@href[1]{\@@startlink{#1}\@@href}%
\providecommand \@@href[1]{\endgroup#1\@@endlink}%
\providecommand \@sanitize@url [0]{\catcode `\\12\catcode `\$12\catcode
  `\&12\catcode `\#12\catcode `\^12\catcode `\_12\catcode `\%12\relax}%
\providecommand \@@startlink[1]{}%
\providecommand \@@endlink[0]{}%
\providecommand \url  [0]{\begingroup\@sanitize@url \@url }%
\providecommand \@url [1]{\endgroup\@href {#1}{\urlprefix }}%
\providecommand \urlprefix  [0]{URL }%
\providecommand \Eprint [0]{\href }%
\providecommand \doibase [0]{http://dx.doi.org/}%
\providecommand \selectlanguage [0]{\@gobble}%
\providecommand \bibinfo  [0]{\@secondoftwo}%
\providecommand \bibfield  [0]{\@secondoftwo}%
\providecommand \translation [1]{[#1]}%
\providecommand \BibitemOpen [0]{}%
\providecommand \bibitemStop [0]{}%
\providecommand \bibitemNoStop [0]{.\EOS\space}%
\providecommand \EOS [0]{\spacefactor3000\relax}%
\providecommand \BibitemShut  [1]{\csname bibitem#1\endcsname}%
\let\auto@bib@innerbib\@empty
\bibitem [{\citenamefont {Kresse}\ and\ \citenamefont
  {Hafner}(1993)}]{PhysRevB.47.558}%
  \BibitemOpen
  \bibfield  {author} {\bibinfo {author} {\bibfnamefont {G.}~\bibnamefont
  {Kresse}}\ and\ \bibinfo {author} {\bibfnamefont {J.}~\bibnamefont
  {Hafner}},\ }\href {\doibase 10.1103/PhysRevB.47.558} {\bibfield  {journal}
  {\bibinfo  {journal} {Phys. Rev. B}\ }\textbf {\bibinfo {volume} {47}},\
  \bibinfo {pages} {558} (\bibinfo {year} {1993})}\BibitemShut {NoStop}%
\bibitem [{\citenamefont {Kresse}\ and\ \citenamefont
  {Furthm\"uller}(1996)}]{PhysRevB.54.11169}%
  \BibitemOpen
  \bibfield  {author} {\bibinfo {author} {\bibfnamefont {G.}~\bibnamefont
  {Kresse}}\ and\ \bibinfo {author} {\bibfnamefont {J.}~\bibnamefont
  {Furthm\"uller}},\ }\href {\doibase 10.1103/PhysRevB.54.11169} {\bibfield
  {journal} {\bibinfo  {journal} {Phys. Rev. B}\ }\textbf {\bibinfo {volume}
  {54}},\ \bibinfo {pages} {11169} (\bibinfo {year} {1996})}\BibitemShut
  {NoStop}%
\bibitem [{\citenamefont {Perdew}\ \emph {et~al.}(1996)\citenamefont {Perdew},
  \citenamefont {Burke},\ and\ \citenamefont
  {Ernzerhof}}]{PhysRevLett.77.3865}%
  \BibitemOpen
  \bibfield  {author} {\bibinfo {author} {\bibfnamefont {J.~P.}\ \bibnamefont
  {Perdew}}, \bibinfo {author} {\bibfnamefont {K.}~\bibnamefont {Burke}}, \
  and\ \bibinfo {author} {\bibfnamefont {M.}~\bibnamefont {Ernzerhof}},\ }\href
  {\doibase 10.1103/PhysRevLett.77.3865} {\bibfield  {journal} {\bibinfo
  {journal} {Phys. Rev. Lett.}\ }\textbf {\bibinfo {volume} {77}},\ \bibinfo
  {pages} {3865} (\bibinfo {year} {1996})}\BibitemShut {NoStop}%
\bibitem [{\citenamefont {Bl\"ochl}(1994)}]{PhysRevB.50.17953}%
  \BibitemOpen
  \bibfield  {author} {\bibinfo {author} {\bibfnamefont {P.~E.}\ \bibnamefont
  {Bl\"ochl}},\ }\href {\doibase 10.1103/PhysRevB.50.17953} {\bibfield
  {journal} {\bibinfo  {journal} {Phys. Rev. B}\ }\textbf {\bibinfo {volume}
  {50}},\ \bibinfo {pages} {17953} (\bibinfo {year} {1994})}\BibitemShut
  {NoStop}%
\bibitem [{\citenamefont {Kresse}\ and\ \citenamefont
  {Joubert}(1999)}]{PhysRevB.59.1758}%
  \BibitemOpen
  \bibfield  {author} {\bibinfo {author} {\bibfnamefont {G.}~\bibnamefont
  {Kresse}}\ and\ \bibinfo {author} {\bibfnamefont {D.}~\bibnamefont
  {Joubert}},\ }\href {\doibase 10.1103/PhysRevB.59.1758} {\bibfield  {journal}
  {\bibinfo  {journal} {Phys. Rev. B}\ }\textbf {\bibinfo {volume} {59}},\
  \bibinfo {pages} {1758} (\bibinfo {year} {1999})}\BibitemShut {NoStop}%
\bibitem [{\citenamefont {Kaltak}\ \emph {et~al.}(2014)\citenamefont {Kaltak},
  \citenamefont {Klime\v{s}},\ and\ \citenamefont
  {Kresse}}]{PhysRevB.90.054115}%
  \BibitemOpen
  \bibfield  {author} {\bibinfo {author} {\bibfnamefont {M.}~\bibnamefont
  {Kaltak}}, \bibinfo {author} {\bibfnamefont {J.}~\bibnamefont {Klime\v{s}}},
  \ and\ \bibinfo {author} {\bibfnamefont {G.}~\bibnamefont {Kresse}},\ }\href
  {\doibase 10.1103/PhysRevB.90.054115} {\bibfield  {journal} {\bibinfo
  {journal} {Phys. Rev. B}\ }\textbf {\bibinfo {volume} {90}},\ \bibinfo
  {pages} {054115} (\bibinfo {year} {2014})}\BibitemShut {NoStop}%
\bibitem [{\citenamefont {Liu}\ \emph {et~al.}(2016)\citenamefont {Liu},
  \citenamefont {Kaltak}, \citenamefont {Klime\v{s}},\ and\ \citenamefont
  {Kresse}}]{PhysRevB.94.165109}%
  \BibitemOpen
  \bibfield  {author} {\bibinfo {author} {\bibfnamefont {P.}~\bibnamefont
  {Liu}}, \bibinfo {author} {\bibfnamefont {M.}~\bibnamefont {Kaltak}},
  \bibinfo {author} {\bibfnamefont {J.}~\bibnamefont {Klime\v{s}}}, \ and\
  \bibinfo {author} {\bibfnamefont {G.}~\bibnamefont {Kresse}},\ }\href
  {\doibase 10.1103/PhysRevB.94.165109} {\bibfield  {journal} {\bibinfo
  {journal} {Phys. Rev. B}\ }\textbf {\bibinfo {volume} {94}},\ \bibinfo
  {pages} {165109} (\bibinfo {year} {2016})}\BibitemShut {NoStop}%
\bibitem [{\citenamefont {Ramberger}\ \emph {et~al.}(2017)\citenamefont
  {Ramberger}, \citenamefont {Sch\"afer},\ and\ \citenamefont
  {Kresse}}]{PhysRevLett.118.106403}%
  \BibitemOpen
  \bibfield  {author} {\bibinfo {author} {\bibfnamefont {B.}~\bibnamefont
  {Ramberger}}, \bibinfo {author} {\bibfnamefont {T.}~\bibnamefont
  {Sch\"afer}}, \ and\ \bibinfo {author} {\bibfnamefont {G.}~\bibnamefont
  {Kresse}},\ }\href {\doibase 10.1103/PhysRevLett.118.106403} {\bibfield
  {journal} {\bibinfo  {journal} {Phys. Rev. Lett.}\ }\textbf {\bibinfo
  {volume} {118}},\ \bibinfo {pages} {106403} (\bibinfo {year}
  {2017})}\BibitemShut {NoStop}%
\bibitem [{\citenamefont {Kaltak}\ and\ \citenamefont
  {Kresse}(2020)}]{PhysRevB.101.205145}%
  \BibitemOpen
  \bibfield  {author} {\bibinfo {author} {\bibfnamefont {M.}~\bibnamefont
  {Kaltak}}\ and\ \bibinfo {author} {\bibfnamefont {G.}~\bibnamefont
  {Kresse}},\ }\href {\doibase 10.1103/PhysRevB.101.205145} {\bibfield
  {journal} {\bibinfo  {journal} {Phys. Rev. B}\ }\textbf {\bibinfo {volume}
  {101}},\ \bibinfo {pages} {205145} (\bibinfo {year} {2020})}\BibitemShut
  {NoStop}%
\bibitem [{\citenamefont {Jinnouchi}\ \emph
  {et~al.}(2019{\natexlab{a}})\citenamefont {Jinnouchi}, \citenamefont
  {Karsai},\ and\ \citenamefont {Kresse}}]{PhysRevB.100.014105}%
  \BibitemOpen
  \bibfield  {author} {\bibinfo {author} {\bibfnamefont {R.}~\bibnamefont
  {Jinnouchi}}, \bibinfo {author} {\bibfnamefont {F.}~\bibnamefont {Karsai}}, \
  and\ \bibinfo {author} {\bibfnamefont {G.}~\bibnamefont {Kresse}},\ }\href
  {\doibase 10.1103/PhysRevB.100.014105} {\bibfield  {journal} {\bibinfo
  {journal} {Phys. Rev. B}\ }\textbf {\bibinfo {volume} {100}},\ \bibinfo
  {pages} {014105} (\bibinfo {year} {2019}{\natexlab{a}})}\BibitemShut
  {NoStop}%
\bibitem [{\citenamefont {Jinnouchi}\ \emph
  {et~al.}(2019{\natexlab{b}})\citenamefont {Jinnouchi}, \citenamefont
  {Lahnsteiner}, \citenamefont {Karsai}, \citenamefont {Kresse},\ and\
  \citenamefont {Bokdam}}]{PhysRevLett.122.225701}%
  \BibitemOpen
  \bibfield  {author} {\bibinfo {author} {\bibfnamefont {R.}~\bibnamefont
  {Jinnouchi}}, \bibinfo {author} {\bibfnamefont {J.}~\bibnamefont
  {Lahnsteiner}}, \bibinfo {author} {\bibfnamefont {F.}~\bibnamefont {Karsai}},
  \bibinfo {author} {\bibfnamefont {G.}~\bibnamefont {Kresse}}, \ and\ \bibinfo
  {author} {\bibfnamefont {M.}~\bibnamefont {Bokdam}},\ }\href {\doibase
  10.1103/PhysRevLett.122.225701} {\bibfield  {journal} {\bibinfo  {journal}
  {Phys. Rev. Lett.}\ }\textbf {\bibinfo {volume} {122}},\ \bibinfo {pages}
  {225701} (\bibinfo {year} {2019}{\natexlab{b}})}\BibitemShut {NoStop}%
\bibitem [{\citenamefont {Liu}\ \emph {et~al.}(2021)\citenamefont {Liu},
  \citenamefont {Verdi}, \citenamefont {Karsai},\ and\ \citenamefont
  {Kresse}}]{Peitao_2020}%
  \BibitemOpen
  \bibfield  {author} {\bibinfo {author} {\bibfnamefont {P.}~\bibnamefont
  {Liu}}, \bibinfo {author} {\bibfnamefont {C.}~\bibnamefont {Verdi}}, \bibinfo
  {author} {\bibfnamefont {F.}~\bibnamefont {Karsai}}, \ and\ \bibinfo {author}
  {\bibfnamefont {G.}~\bibnamefont {Kresse}},\ }\href {\doibase
  10.1103/PhysRevMaterials.5.053804} {\bibfield  {journal} {\bibinfo  {journal}
  {Phys. Rev. Materials}\ }\textbf {\bibinfo {volume} {5}},\ \bibinfo {pages}
  {053804} (\bibinfo {year} {2021})}\BibitemShut {NoStop}%
\bibitem [{\citenamefont {Verdi}\ \emph {et~al.}(2021)\citenamefont {Verdi},
  \citenamefont {Karsai}, \citenamefont {Liu}, \citenamefont {Jinnouchi},\ and\
  \citenamefont {Kresse}}]{Carla_2021}%
  \BibitemOpen
  \bibfield  {author} {\bibinfo {author} {\bibfnamefont {C.}~\bibnamefont
  {Verdi}}, \bibinfo {author} {\bibfnamefont {F.}~\bibnamefont {Karsai}},
  \bibinfo {author} {\bibfnamefont {P.}~\bibnamefont {Liu}}, \bibinfo {author}
  {\bibfnamefont {R.}~\bibnamefont {Jinnouchi}}, \ and\ \bibinfo {author}
  {\bibfnamefont {G.}~\bibnamefont {Kresse}},\ }\href {\doibase
  10.1038/s41524-021-00630-5} {\bibfield  {journal} {\bibinfo  {journal} {npj
  Computational Materials}\ }\textbf {\bibinfo {volume} {7}},\ \bibinfo {pages}
  {156} (\bibinfo {year} {2021})}\BibitemShut {NoStop}%
\bibitem [{\citenamefont {Jinnouchi}\ \emph
  {et~al.}(2020{\natexlab{a}})\citenamefont {Jinnouchi}, \citenamefont
  {Karsai},\ and\ \citenamefont {Kresse}}]{PhysRevB.101.060201}%
  \BibitemOpen
  \bibfield  {author} {\bibinfo {author} {\bibfnamefont {R.}~\bibnamefont
  {Jinnouchi}}, \bibinfo {author} {\bibfnamefont {F.}~\bibnamefont {Karsai}}, \
  and\ \bibinfo {author} {\bibfnamefont {G.}~\bibnamefont {Kresse}},\ }\href
  {\doibase 10.1103/PhysRevB.101.060201} {\bibfield  {journal} {\bibinfo
  {journal} {Phys. Rev. B}\ }\textbf {\bibinfo {volume} {101}},\ \bibinfo
  {pages} {060201} (\bibinfo {year} {2020}{\natexlab{a}})}\BibitemShut
  {NoStop}%
\bibitem [{\citenamefont {Allen}\ and\ \citenamefont
  {Tildesley}(1991)}]{book_Allen_Tildesley}%
  \BibitemOpen
  \bibfield  {author} {\bibinfo {author} {\bibfnamefont {M.~P.}\ \bibnamefont
  {Allen}}\ and\ \bibinfo {author} {\bibfnamefont {D.~J.}\ \bibnamefont
  {Tildesley}},\ }\href@noop {} {\emph {\bibinfo {title} {Computer simulation
  of liquids (Oxford university press: New York)}}}\ (\bibinfo {year}
  {1991})\BibitemShut {NoStop}%
\bibitem [{\citenamefont {Parrinello}\ and\ \citenamefont
  {Rahman}(1980)}]{PhysRevLett.45.1196}%
  \BibitemOpen
  \bibfield  {author} {\bibinfo {author} {\bibfnamefont {M.}~\bibnamefont
  {Parrinello}}\ and\ \bibinfo {author} {\bibfnamefont {A.}~\bibnamefont
  {Rahman}},\ }\href {\doibase 10.1103/PhysRevLett.45.1196} {\bibfield
  {journal} {\bibinfo  {journal} {Phys. Rev. Lett.}\ }\textbf {\bibinfo
  {volume} {45}},\ \bibinfo {pages} {1196} (\bibinfo {year}
  {1980})}\BibitemShut {NoStop}%
\bibitem [{\citenamefont {Jinnouchi}\ \emph
  {et~al.}(2020{\natexlab{b}})\citenamefont {Jinnouchi}, \citenamefont
  {Karsai}, \citenamefont {Verdi}, \citenamefont {Asahi},\ and\ \citenamefont
  {Kresse}}]{doi:10.1063/5.0009491}%
  \BibitemOpen
  \bibfield  {author} {\bibinfo {author} {\bibfnamefont {R.}~\bibnamefont
  {Jinnouchi}}, \bibinfo {author} {\bibfnamefont {F.}~\bibnamefont {Karsai}},
  \bibinfo {author} {\bibfnamefont {C.}~\bibnamefont {Verdi}}, \bibinfo
  {author} {\bibfnamefont {R.}~\bibnamefont {Asahi}}, \ and\ \bibinfo {author}
  {\bibfnamefont {G.}~\bibnamefont {Kresse}},\ }\href {\doibase
  10.1063/5.0009491} {\bibfield  {journal} {\bibinfo  {journal} {J. Chem.
  Phys.}\ }\textbf {\bibinfo {volume} {152}},\ \bibinfo {pages} {234102}
  (\bibinfo {year} {2020}{\natexlab{b}})}\BibitemShut {NoStop}%
\bibitem [{\citenamefont {Bart\'ok}\ \emph {et~al.}(2013)\citenamefont
  {Bart\'ok}, \citenamefont {Kondor},\ and\ \citenamefont
  {Cs\'anyi}}]{PhysRevB.87.184115}%
  \BibitemOpen
  \bibfield  {author} {\bibinfo {author} {\bibfnamefont {A.~P.}\ \bibnamefont
  {Bart\'ok}}, \bibinfo {author} {\bibfnamefont {R.}~\bibnamefont {Kondor}}, \
  and\ \bibinfo {author} {\bibfnamefont {G.}~\bibnamefont {Cs\'anyi}},\ }\href
  {\doibase 10.1103/PhysRevB.87.184115} {\bibfield  {journal} {\bibinfo
  {journal} {Phys. Rev. B}\ }\textbf {\bibinfo {volume} {87}},\ \bibinfo
  {pages} {184115} (\bibinfo {year} {2013})}\BibitemShut {NoStop}%
\bibitem [{\citenamefont {Liu}\ \emph {et~al.}(2022)\citenamefont {Liu},
  \citenamefont {Verdi}, \citenamefont {Karsai},\ and\ \citenamefont
  {Kresse}}]{PhysRevB.105.L060102}%
  \BibitemOpen
  \bibfield  {author} {\bibinfo {author} {\bibfnamefont {P.}~\bibnamefont
  {Liu}}, \bibinfo {author} {\bibfnamefont {C.}~\bibnamefont {Verdi}}, \bibinfo
  {author} {\bibfnamefont {F.}~\bibnamefont {Karsai}}, \ and\ \bibinfo {author}
  {\bibfnamefont {G.}~\bibnamefont {Kresse}},\ }\href {\doibase
  10.1103/PhysRevB.105.L060102} {\bibfield  {journal} {\bibinfo  {journal}
  {Phys. Rev. B}\ }\textbf {\bibinfo {volume} {105}},\ \bibinfo {pages}
  {L060102} (\bibinfo {year} {2022})}\BibitemShut {NoStop}%
\bibitem [{\citenamefont {Ramakrishnan}\ \emph {et~al.}(2015)\citenamefont
  {Ramakrishnan}, \citenamefont {Dral}, \citenamefont {Rupp},\ and\
  \citenamefont {von Lilienfeld}}]{Anatole_2015JCTC}%
  \BibitemOpen
  \bibfield  {author} {\bibinfo {author} {\bibfnamefont {R.}~\bibnamefont
  {Ramakrishnan}}, \bibinfo {author} {\bibfnamefont {P.~O.}\ \bibnamefont
  {Dral}}, \bibinfo {author} {\bibfnamefont {M.}~\bibnamefont {Rupp}}, \ and\
  \bibinfo {author} {\bibfnamefont {O.~A.}\ \bibnamefont {von Lilienfeld}},\
  }\href {\doibase 10.1021/acs.jctc.5b00099} {\bibfield  {journal} {\bibinfo
  {journal} {J. Chem. Theory Comput.}\ }\textbf {\bibinfo {volume} {11}},\
  \bibinfo {pages} {2087} (\bibinfo {year} {2015})}\BibitemShut {NoStop}%
\bibitem [{\citenamefont {Cheng}\ \emph {et~al.}(2020)\citenamefont {Cheng},
  \citenamefont {Griffiths}, \citenamefont {Wengert}, \citenamefont {Kunkel},
  \citenamefont {Stenczel}, \citenamefont {Zhu}, \citenamefont {Deringer},
  \citenamefont {Bernstein}, \citenamefont {Margraf}, \citenamefont {Reuter},\
  and\ \citenamefont {Csanyi}}]{doi:10.1021/acs.accounts.0c00403}%
  \BibitemOpen
  \bibfield  {author} {\bibinfo {author} {\bibfnamefont {B.}~\bibnamefont
  {Cheng}}, \bibinfo {author} {\bibfnamefont {R.-R.}\ \bibnamefont
  {Griffiths}}, \bibinfo {author} {\bibfnamefont {S.}~\bibnamefont {Wengert}},
  \bibinfo {author} {\bibfnamefont {C.}~\bibnamefont {Kunkel}}, \bibinfo
  {author} {\bibfnamefont {T.}~\bibnamefont {Stenczel}}, \bibinfo {author}
  {\bibfnamefont {B.}~\bibnamefont {Zhu}}, \bibinfo {author} {\bibfnamefont
  {V.~L.}\ \bibnamefont {Deringer}}, \bibinfo {author} {\bibfnamefont
  {N.}~\bibnamefont {Bernstein}}, \bibinfo {author} {\bibfnamefont {J.~T.}\
  \bibnamefont {Margraf}}, \bibinfo {author} {\bibfnamefont {K.}~\bibnamefont
  {Reuter}}, \ and\ \bibinfo {author} {\bibfnamefont {G.}~\bibnamefont
  {Csanyi}},\ }\href {\doibase 10.1021/acs.accounts.0c00403} {\bibfield
  {journal} {\bibinfo  {journal} {Accounts of Chemical Research}\ }\textbf
  {\bibinfo {volume} {53}},\ \bibinfo {pages} {1981} (\bibinfo {year}
  {2020})}\BibitemShut {NoStop}%
\bibitem [{\citenamefont {Shapeev}(2016)}]{Shapeev_MTP2016}%
  \BibitemOpen
  \bibfield  {author} {\bibinfo {author} {\bibfnamefont {A.~V.}\ \bibnamefont
  {Shapeev}},\ }\href {\doibase 10.1137/15M1054183} {\bibfield  {journal}
  {\bibinfo  {journal} {Multiscale Modeling \& Simulation}\ }\textbf {\bibinfo
  {volume} {14}},\ \bibinfo {pages} {1153} (\bibinfo {year}
  {2016})}\BibitemShut {NoStop}%
\bibitem [{\citenamefont {Novikov}\ \emph {et~al.}(2021)\citenamefont
  {Novikov}, \citenamefont {Gubaev}, \citenamefont {Podryabinkin},\ and\
  \citenamefont {Shapeev}}]{Novikov_2021}%
  \BibitemOpen
  \bibfield  {author} {\bibinfo {author} {\bibfnamefont {I.~S.}\ \bibnamefont
  {Novikov}}, \bibinfo {author} {\bibfnamefont {K.}~\bibnamefont {Gubaev}},
  \bibinfo {author} {\bibfnamefont {E.~V.}\ \bibnamefont {Podryabinkin}}, \
  and\ \bibinfo {author} {\bibfnamefont {A.~V.}\ \bibnamefont {Shapeev}},\
  }\href {\doibase 10.1088/2632-2153/abc9fe} {\bibfield  {journal} {\bibinfo
  {journal} {Machine Learning: Science and Technology}\ }\textbf {\bibinfo
  {volume} {2}},\ \bibinfo {pages} {025002} (\bibinfo {year}
  {2021})}\BibitemShut {NoStop}%
\bibitem [{\citenamefont {Lazi\ifmmode~\acute{c}\else \'{c}\fi{}}\ \emph
  {et~al.}(2010)\citenamefont {Lazi\ifmmode~\acute{c}\else \'{c}\fi{}},
  \citenamefont {Alaei}, \citenamefont {Atodiresei}, \citenamefont {Caciuc},
  \citenamefont {Brako},\ and\ \citenamefont {Bl\"ugel}}]{PhysRevB.81.045401}%
  \BibitemOpen
  \bibfield  {author} {\bibinfo {author} {\bibfnamefont {P.}~\bibnamefont
  {Lazi\ifmmode~\acute{c}\else \'{c}\fi{}}}, \bibinfo {author} {\bibfnamefont
  {M.}~\bibnamefont {Alaei}}, \bibinfo {author} {\bibfnamefont
  {N.}~\bibnamefont {Atodiresei}}, \bibinfo {author} {\bibfnamefont
  {V.}~\bibnamefont {Caciuc}}, \bibinfo {author} {\bibfnamefont
  {R.}~\bibnamefont {Brako}}, \ and\ \bibinfo {author} {\bibfnamefont
  {S.}~\bibnamefont {Bl\"ugel}},\ }\href {\doibase 10.1103/PhysRevB.81.045401}
  {\bibfield  {journal} {\bibinfo  {journal} {Phys. Rev. B}\ }\textbf {\bibinfo
  {volume} {81}},\ \bibinfo {pages} {045401} (\bibinfo {year}
  {2010})}\BibitemShut {NoStop}%
\bibitem [{\citenamefont {Dion}\ \emph {et~al.}(2004)\citenamefont {Dion},
  \citenamefont {Rydberg}, \citenamefont {Schr\"oder}, \citenamefont
  {Langreth},\ and\ \citenamefont {Lundqvist}}]{PhysRevLett.92.246401}%
  \BibitemOpen
  \bibfield  {author} {\bibinfo {author} {\bibfnamefont {M.}~\bibnamefont
  {Dion}}, \bibinfo {author} {\bibfnamefont {H.}~\bibnamefont {Rydberg}},
  \bibinfo {author} {\bibfnamefont {E.}~\bibnamefont {Schr\"oder}}, \bibinfo
  {author} {\bibfnamefont {D.~C.}\ \bibnamefont {Langreth}}, \ and\ \bibinfo
  {author} {\bibfnamefont {B.~I.}\ \bibnamefont {Lundqvist}},\ }\href {\doibase
  10.1103/PhysRevLett.92.246401} {\bibfield  {journal} {\bibinfo  {journal}
  {Phys. Rev. Lett.}\ }\textbf {\bibinfo {volume} {92}},\ \bibinfo {pages}
  {246401} (\bibinfo {year} {2004})}\BibitemShut {NoStop}%
\bibitem [{\citenamefont {Grimme}\ \emph {et~al.}(2010)\citenamefont {Grimme},
  \citenamefont {Antony}, \citenamefont {Ehrlich},\ and\ \citenamefont
  {Krieg}}]{doi:10.1063/1.3382344}%
  \BibitemOpen
  \bibfield  {author} {\bibinfo {author} {\bibfnamefont {S.}~\bibnamefont
  {Grimme}}, \bibinfo {author} {\bibfnamefont {J.}~\bibnamefont {Antony}},
  \bibinfo {author} {\bibfnamefont {S.}~\bibnamefont {Ehrlich}}, \ and\
  \bibinfo {author} {\bibfnamefont {H.}~\bibnamefont {Krieg}},\ }\href
  {\doibase 10.1063/1.3382344} {\bibfield  {journal} {\bibinfo  {journal} {The
  Journal of Chemical Physics}\ }\textbf {\bibinfo {volume} {132}},\ \bibinfo
  {pages} {154104} (\bibinfo {year} {2010})}\BibitemShut {NoStop}%
\bibitem [{\citenamefont {Hammer}\ \emph {et~al.}(1999)\citenamefont {Hammer},
  \citenamefont {Hansen},\ and\ \citenamefont {N\o{}rskov}}]{PhysRevB.59.7413}%
  \BibitemOpen
  \bibfield  {author} {\bibinfo {author} {\bibfnamefont {B.}~\bibnamefont
  {Hammer}}, \bibinfo {author} {\bibfnamefont {L.~B.}\ \bibnamefont {Hansen}},
  \ and\ \bibinfo {author} {\bibfnamefont {J.~K.}\ \bibnamefont {N\o{}rskov}},\
  }\href {\doibase 10.1103/PhysRevB.59.7413} {\bibfield  {journal} {\bibinfo
  {journal} {Phys. Rev. B}\ }\textbf {\bibinfo {volume} {59}},\ \bibinfo
  {pages} {7413} (\bibinfo {year} {1999})}\BibitemShut {NoStop}%
\bibitem [{\citenamefont {Tyson}\ and\ \citenamefont
  {Miller}(1977)}]{TYSON1977267}%
  \BibitemOpen
  \bibfield  {author} {\bibinfo {author} {\bibfnamefont {W.}~\bibnamefont
  {Tyson}}\ and\ \bibinfo {author} {\bibfnamefont {W.}~\bibnamefont {Miller}},\
  }\href {\doibase https://doi.org/10.1016/0039-6028(77)90442-3} {\bibfield
  {journal} {\bibinfo  {journal} {Surface Science}\ }\textbf {\bibinfo {volume}
  {62}},\ \bibinfo {pages} {267} (\bibinfo {year} {1977})}\BibitemShut
  {NoStop}%
\bibitem [{\citenamefont {Abild-Pedersen}\ and\ \citenamefont
  {Andersson}(2007)}]{ABILDPEDERSEN20071747}%
  \BibitemOpen
  \bibfield  {author} {\bibinfo {author} {\bibfnamefont {F.}~\bibnamefont
  {Abild-Pedersen}}\ and\ \bibinfo {author} {\bibfnamefont {M.}~\bibnamefont
  {Andersson}},\ }\href {\doibase https://doi.org/10.1016/j.susc.2007.01.052}
  {\bibfield  {journal} {\bibinfo  {journal} {Surface Science}\ }\textbf
  {\bibinfo {volume} {601}},\ \bibinfo {pages} {1747} (\bibinfo {year}
  {2007})}\BibitemShut {NoStop}%
\bibitem [{\citenamefont {Linke}\ \emph {et~al.}(2001)\citenamefont {Linke},
  \citenamefont {Curulla}, \citenamefont {Hopstaken},\ and\ \citenamefont
  {Niemantsverdriet}}]{doi:10.1063/1.1355767}%
  \BibitemOpen
  \bibfield  {author} {\bibinfo {author} {\bibfnamefont {R.}~\bibnamefont
  {Linke}}, \bibinfo {author} {\bibfnamefont {D.}~\bibnamefont {Curulla}},
  \bibinfo {author} {\bibfnamefont {M.~J.~P.}\ \bibnamefont {Hopstaken}}, \
  and\ \bibinfo {author} {\bibfnamefont {J.~W.}\ \bibnamefont
  {Niemantsverdriet}},\ }\href {\doibase 10.1063/1.1355767} {\bibfield
  {journal} {\bibinfo  {journal} {The Journal of Chemical Physics}\ }\textbf
  {\bibinfo {volume} {115}},\ \bibinfo {pages} {8209} (\bibinfo {year}
  {2001})}\BibitemShut {NoStop}%
\end{thebibliography}%

\end{document}


\title{Supplementary Material to \\
``Combining Machine Learning and Many-Body Calculations: \\
Coverage-Dependent Adsorption of CO on Rh(111)"}

\author{Peitao Liu}
\email{ptliu@imr.ac.cn}
\affiliation{University of Vienna, Faculty of Physics and Center for Computational Materials Science, Kolingasse 14-16, A-1090 Vienna, Austria}
\affiliation{Shenyang National Laboratory for Materials Science, Institute of Metal Research, Chinese Academy of Sciences, 110016 Shenyang, China}

\author{Jiantao Wang}
\affiliation{Shenyang National Laboratory for Materials Science, Institute of Metal Research, Chinese Academy of Sciences, 110016 Shenyang, China}

\author{Noah Avargues}
\affiliation{University of Vienna, Faculty of Physics and Center for Computational Materials Science, Kolingasse 14-16, A-1090 Vienna, Austria}

\author{Carla Verdi}
\affiliation{University of Vienna, Faculty of Physics and Center for Computational Materials Science, Kolingasse 14-16, A-1090 Vienna, Austria}

\author{Andreas Singraber}
\affiliation{VASP Software GmbH, Sensengasse 8, A-1090 Vienna, Austria}

\author{Ferenc Karsai}
\affiliation{VASP Software GmbH, Sensengasse 8, A-1090 Vienna, Austria}

\author{Xing-Qiu Chen}
\affiliation{Shenyang National Laboratory for Materials Science, Institute of Metal Research, Chinese Academy of Sciences, 110016 Shenyang, China}

\author{Georg Kresse}
\affiliation{University of Vienna, Faculty of Physics and Center for Computational Materials Science, Kolingasse 14-16, A-1090 Vienna, Austria}
\affiliation{VASP Software GmbH, Sensengasse 8, A-1090 Vienna, Austria}

\maketitle

\section{First-principles calculations}\label{sec:DFT_details}

The first-principles (FP) calculations were performed using the Vienna
\emph{Ab initio} Simulation Package (VASP)~\cite{PhysRevB.47.558, PhysRevB.54.11169}.
The plane-wave cutoff for the orbitals was chosen to be 450 eV.
A $\Gamma$-centered $k$-point grid with a spacing of 0.2 $\AA^{-1}$
between $k$ points was used to sample the Brillouin zone.
For density functional theory (DFT) calculations,
the electronic interactions were described using the Perdew-Burke-Ernzerhof (PBE) functional~\cite{PhysRevLett.77.3865}.
The projector augmented wave (PAW) pseudopotentials~\cite{PhysRevB.50.17953,PhysRevB.59.1758} ({\tt Rh}, {\tt C} and {\tt O})
with the valence electron configurations of $4d^{8}5s^1$, $2s^22p^2$, and $2s^22p^4$ were employed for Rh, C, and O, respectively.
The Gaussian smearing method with a smearing width of 0.05 eV was used.
Whenever ground-state structures were required, the electronic optimization was performed until the total energy difference between two iterations was less than 10$^{-5}$ eV.
The structures were optimized until the forces were smaller than 0.01 eV/\AA.

For the random phase approximation (RPA) calculations, the GW PAW
potentials ({\tt Rh\_sv\_GW}, {\tt C\_GW} and {\tt O\_GW}) were used.
The energy cutoff for the response function was chosen to be 300 eV.
The RPA energies and forces were calculated using an efficient cubic scaling
algorithm~\cite{PhysRevB.90.054115,PhysRevB.94.165109,PhysRevLett.118.106403}.
To handle the partial occupancies of states at the Fermi level for metallic systems,
the finite-temperature RPA algorithm~\cite{PhysRevB.101.205145} was used.
This algorithm employs a compressive sensing approach for the Matsubara summation so that
a small number of imaginary frequency/time points is sufficient to obtain a reasonably high accuracy~\cite{PhysRevB.101.205145}.
Here, the number of imaginary frequency/time points was set to 16 and the Fermi smearing method with a smearing width of 0.15 eV was chosen after
extensive convergence tests.

\section{MLFF training}\label{sec:training_details}

Here, we give additional details on the machine-learned forced fields (MLFFs) training, while most of the training procedure is outlined in the main text.
Our training structures were automatically selected during FP molecular dynamics (MD) simulations
by the on-the-fly active learning method based on the Bayesian linear regression (BLR)~\cite{PhysRevB.100.014105,PhysRevLett.122.225701}.
The on-the-fly learning method has been successfully used to develop accurate MLFFs for the prediction of
various phase transitions~\cite{PhysRevLett.122.225701,Peitao_2020,Carla_2021},
melting points~\cite{PhysRevB.100.014105}, lattice thermal conductivities~\cite{Carla_2021}
as well as chemical potentials~\cite{PhysRevB.101.060201}.
The FPMD simulations were performed at ambient pressure using a Langevin thermostat~\cite{book_Allen_Tildesley}
combined with the Parrinello-Raman method~\cite{PhysRevLett.45.1196}.
The isothermal-isobaric (NPT) and the canonical (NVT)  ensembles were used for bulk Rh and slab calculations, respectively.
The PBE functional~\cite{PhysRevLett.77.3865} and a time step of 2 fs were employed.
For the on-the-fly learning, the atom density based separable descriptors~\cite{doi:10.1063/5.0009491} were used.
The separable descriptors are similar to the smooth overlap of atomic positions (SOAP) formulation
proposed by Bart\'{o}k \emph{et al.}~\cite{PhysRevB.87.184115}, but differ in that
the two-body distribution functions are explicitly included and
the three-body distribution functions are free of two-body components by subtracting the self-interaction term
so that the weight of the two- and three-body descriptors can be separately tuned~\cite{doi:10.1063/5.0009491}.
For more discussions on the difference between the separable descriptors and the SOAP descriptors, we refer to Ref.~\cite{doi:10.1063/5.0009491}.
The cutoff radius was set to 5 $\AA$ for both two- and three-body descriptors.
The width of the Gaussian functions used for broadening the atomic distributions
of the two- and three-body descriptors were set to 0.27 and 0.40, respectively.
The number of radial basis functions were set to 21 and 14 for radial and angular parts, respectively.

\begin{table}
\caption {Summary of the structures included in the training dataset $T^A$.
The employed slab models here consist of 6 Rh layers and a vacuum width of 15~$\AA$.}
\begin{ruledtabular}
\begin{tabular}{crr}
 CO coverages (ML) & Structure type                  & No. structures \\
\hline
 ---  &  Bulk Rh of 4-atom cell         &  115  \\
 ---  &  Bulk Rh of 32-atom cell        &   85  \\
 0/4  &  Clean 2x2Rh(111)               &  196  \\
 1/4  &  1CO@2x2Rh(111)                 & 1327  \\
 2/4  &  2CO@2x2Rh(111)                 & 1361  \\
 3/4  &  3CO@2x2Rh(111)                 &  294  \\
 1/16  &  1CO@4x4Rh(111)                 &  116  \\
 2/16  &  2CO@4x4Rh(111)                 &   203  \\
 3/16  &  3CO@4x4Rh(111)                 &   252  \\
4/16  &  4CO@4x4Rh(111)                 &  240  \\
5/16  &  5CO@4x4Rh(111)                 &   67  \\
6/16  &  6CO@4x4Rh(111)                 &  148  \\
7/16  &  7CO@4x4Rh(111)                 &   85  \\
8/16  &  8CO@4x4Rh(111)                 &  161  \\
9/16  &  9CO@4x4Rh(111)                 &   66  \\
10/16  & 10CO@4x4Rh(111)                 &  144  \\
11/16  & 11CO@4x4Rh(111)                 &   85  \\
12/16  & 12CO@4x4Rh(111)                 &  177  \\
2/9  &  2CO@3x3Rh(111)                 &   11  \\
3/9  &  3CO@3x3Rh(111)                 &   15  \\
3/36  &  3CO@6x6Rh(111)                 &   28  \\
4/36  &  4CO@6x6Rh(111)                 &   18  \\
6/36  &  6CO@6x6Rh(111)                 &   22  \\
9/36  &  9CO@6x6Rh(111)                 &   17  \\
\hline
Total    &                                    & 5267 \\
\end{tabular}
\end{ruledtabular}
\label{tab:training_dataset_A}
\end{table}

\begin{table}
\caption {Summary of the structures included in the training dataset $T^B$.
The employed slab models here consist of 4 Rh layers and a vacuum width of 10~$\AA$}
\begin{ruledtabular}
\begin{tabular}{crr}
CO coverages (ML)  & Structure type                  & No. structures \\
\hline
---  &  Bulk Rh of 32-atom cell        &   2  \\
0/4  &  Clean 2x2Rh(111)               &  1  \\
1/4  &  1CO@2x2Rh(111)                 & 18  \\
2/4  &  2CO@2x2Rh(111)                 & 24  \\
3/4  &  3CO@2x2Rh(111)                 &  55 \\
\hline
Total    &                                    & 100  \\
\end{tabular}
\end{ruledtabular}
\label{tab:training_dataset_B}
\end{table}

\begin{figure}
\begin{center}
\includegraphics[width=0.6\textwidth,trim = {0.0cm 0.0cm 0.0cm 0.0cm}, clip]{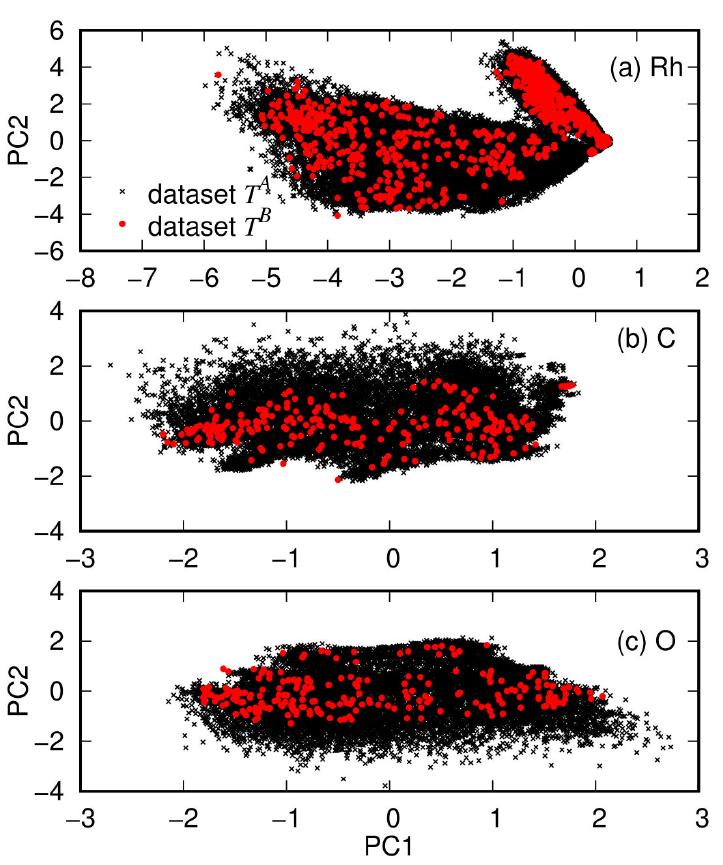}
\end{center}
\caption{KPCA map for the local atomic environments of (a) Rh, (b) C, and (c) O atoms in the datasets  $T^A$ (black star) and $T^B$ (red circle).
}
\label{fig:PCA}
\end{figure}

We followed the method proposed in our recent work of Ref.~\cite{PhysRevB.105.L060102} to develop an RPA-derived MLFF (MLFF-RPA$^\Delta$)
for the prediction of coverage-dependent adsorption of CO on the Rh(111) surface.
As explained in the main text, the development of MLFF-RPA$^\Delta$ involves the construction of a baseline PBE-derived MLFF (MLFF-PBE)
and an intermediate surrogate MLFF model for machine-learning the differences
between RPA and PBE calculated energies and forces (called MLFF-$\Delta$ where $\Delta$=RPA$-$PBE).
The success of the $\Delta$-ML approach originates from the  assumption that low-level reference quantum-mechanical (QM) calculations
such as DFT capture the most important contributions to the overall potential energy surface (though they might not be accurate)
and therefore the remaining differences between high-level and low-level QM calculations become less corrugated
and thus easier to be machine-learned~\cite{Anatole_2015JCTC}.
The first training dataset ($T^A$) contains 5267 structures,
whose detailed information is summarized in Table~\ref{tab:training_dataset_A}.
The second dataset referred to as $T^B$ contains 100 structures, which are detailed in Table~\ref{tab:training_dataset_B}.
Through the kernel principal component analysis (KPCA)~\cite{doi:10.1021/acs.accounts.0c00403}
on local atomic environments of each species, we found that these 100 structures of small unit cells in $T^B$
are indeed very representative and cover a wide range of the phase space, as shown in Fig.~\ref{fig:PCA}.
For the structures in  the dataset $T^B$, the energies and forces were recalculated using finite-temperature RPA calculations.
The MLFF-$\Delta$ was generated by machine-learning the differences in energies and forces
between RPA and PBE calculations using the separable descriptors with a low spatial resolution of 0.6~$\AA$
and a small number of radial basis functions of 10 for both the radial and angular parts.
Finally, the energies and forces of the structures in $T^A$
were corrected by adding the differences predicted by the MLFF-$\Delta$.
The final MLFF-RPA$^\Delta$ was then fitted to the updated $T^A$.

\begin{table}
\caption {Summary of the structures included in the validation dataset $T^C$.}
\begin{ruledtabular}
\begin{tabular}{crr}
CO coverages (ML)  & Structure type                  & No. structures \\
\hline
---  &  Bulk Rh of 32-atom cell        &   40  \\
0/4 &  Clean 2x2Rh(111)               &  40  \\
2/16  &  2CO@4x4Rh(111)                 & 40  \\
1/4  &  1CO@2x2Rh(111)                 & 40  \\
6/16  &  6CO@4x4Rh(111)                 & 40  \\
2/4  &  2CO@2x2Rh(111)                 & 40  \\
10/16  &  10CO@4x4Rh(111)                 & 40  \\
3/4  &  3CO@2x2Rh(111)                 & 40  \\
\hline
Total    &                                    & 320  \\
\end{tabular}
\end{ruledtabular}
\label{tab:test_dataset_C}
\end{table}

We note that the final regression coefficients were determined by using a factor of 20
for the relative weight of the energy equations with respect to the equations for the forces
and a singular value decomposition (SVD) to solve the least-squares problem. This was found to improve the overall accuracy
of MLFFs as compared to BLR~\cite{Peitao_2020,Carla_2021,PhysRevB.105.L060102}.
As explained in the main text, we also constructed moment tensor potentials (MTPs)~\cite{Shapeev_MTP2016} for both PBE and RPA
using the MLIP package~\cite{Novikov_2021}
based on the same training datasets as used in generating the kernel-based MLFFs.
The MTP basis functions (i.e., contractions of one or more moments) were selected such that
the level of scalar basis $B_\alpha$ is less than or equal to 26 (i.e, $\text{lev} B_{\alpha} \leq 26$).
The number of radial basis functions was chosen to be 7.
The regression coefficients were obtained
by a non-linear least square optimization using the Broyden-Fletcher-Goldfarb-Shanno (BFGS) algorithm.
The weights expressing the importance of energies and forces in the optimization were set to be 1 and 0.025, respectively,
which were found to yield the overall lowest loss function.
To obtain more accurate MLFFs, a larger cutoff radius of 6~$\AA$ was finally used for retraining.
It needs to be noted that a cutoff radius of 6~$\AA$ for a many-body potential
allows to model interactions up to twice that cutoff, i.e. 12~$\AA$.
This suffices to capture all the relevant van der Waals (vdW) interactions.
Moreover, it has been shown for CO on Pt(111) that nonlocal correlations beyond 4~$\AA$
are not relevant for predicting adsorption energy differences~\cite{PhysRevB.81.045401}.
To avoid confusions, the shorthand ``MTP'' always refers to the moment tensor potential and ``KMLFF" to the kernel-based approach used in VASP.

We also note that the isolated CO molecule was not learned by MLFFs.
Instead, we calculated its total energy ($E_{\rm CO}$) using PBE or RPA
and fixed these values throughout all the CO adsorption energy calculations.
The adsorption energy per CO [$E_{\rm ad}(n)$] for $n$ CO molecules adsorbed on the surface is calculated as
\begin{equation}
E_{\rm ad}(n) = \frac{1}{n}(E_{\rm total} - E_{\rm slab} - nE_{\rm CO}),
\end{equation}
where $E_{\rm total}$ and $E_{\rm slab}$ are the total energies of the optimized CO@Rh(111) complex and the clean Rh(111) surface, respectively.
The total binding energy of the system is then obtained by
\begin{equation}
E_{\rm b}(n) = n E_{\rm ad}(n).
\end{equation}

\section{MLFF validation}\label{sec:validation_details}

\begin{table}
\caption {A summary of structures included in the validation dataset $T^D$.}
\begin{ruledtabular}
\begin{tabular}{crr}
CO coverages (ML)  & Structure type                  & No. structures \\
\hline
0/4 &  Clean 2x2Rh(111)               &  1  \\
1/4  &  1CO@2x2Rh(111)                 & 8  \\
2/4  &  2CO@2x2Rh(111)                 & 8  \\
3/4  &  3CO@2x2Rh(111)                 & 8  \\
\hline
Total    &                                    & 25  \\
\end{tabular}
\end{ruledtabular}
\label{tab:test_dataset_D}
\end{table}

\begin{figure}
\begin{center}
\includegraphics[width=0.6\textwidth,trim = {0.0cm 0.0cm 0.0cm 0.0cm}, clip]{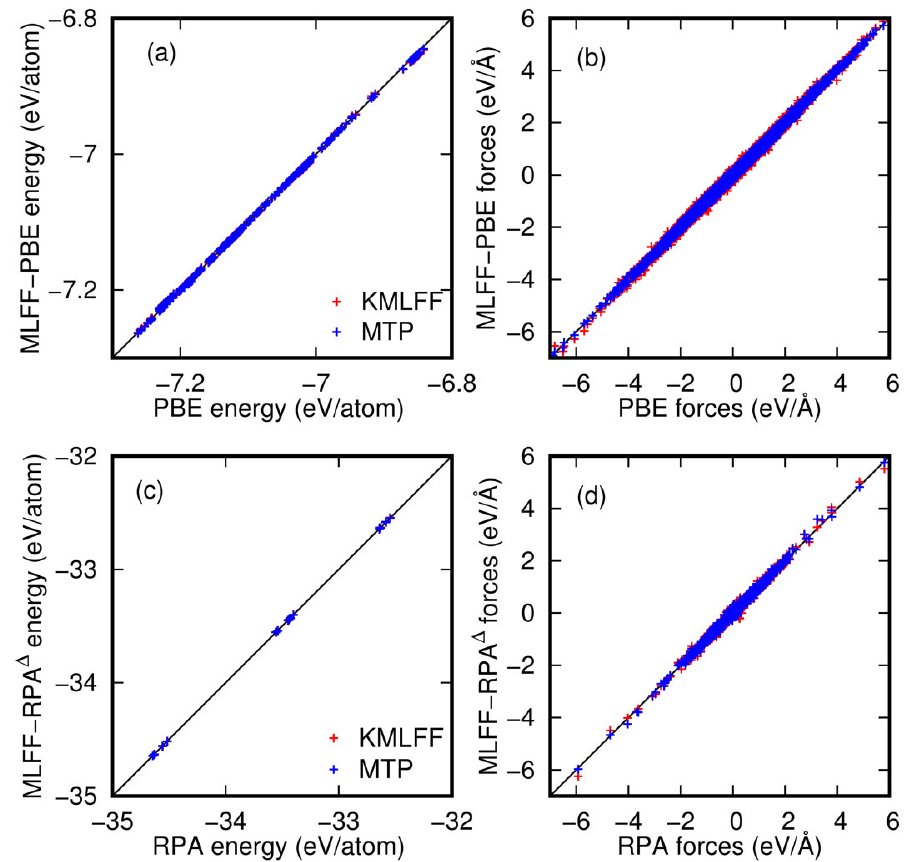}
\end{center}
 \caption{(a, b) MLFFs-PBE vs. PBE in energies and forces for the validation dataset $T^C$ containing 320 structures (see Table~\ref{tab:test_dataset_C}).
 (c, d) MLFFs-RPA$^\Delta$ vs. RPA in energies and forces for the validation dataset $T^D$ containing 25 structures (see Table~\ref{tab:test_dataset_D}).
 The results obtained from KMLFFs and MTPs are shown in red and blue, respectively.
}
\label{fig:Error_analysis}
\end{figure}

The MLFFs-PBE were validated on a test dataset ($T^C$) containing 320 structures
including bulk Rh, the clean Rh(111) surface, and CO adsorption on the Rh(111) surface with various coverages.
These structures were randomly selected from MD simulations at $T$=1000 K using the MTP-RPA$^\Delta$. The detailed information
for the test dataset $T^C$ is summarized in Table~\ref{tab:test_dataset_C}.
Due to the large computational cost of RPA calculations,
the validation of MLFFs-RPA$^\Delta$ was restricted to a reduced test dataset ($T^D$) of small unit cells (see Table~\ref{tab:test_dataset_D}).
The validation root-mean-square errors (RMSEs) in energies and forces for MLFFs-PBE and MLFFs-RPA$^\Delta$
using atom density based descriptors or moment tensor descriptors are given in Table I of the main text
and the MLFFs predicted energies and forces against the first-principles data are further illustrated in Fig.~\ref{fig:Error_analysis}.

\begin{figure}
\begin{center}
\includegraphics[width=0.45\textwidth, clip]{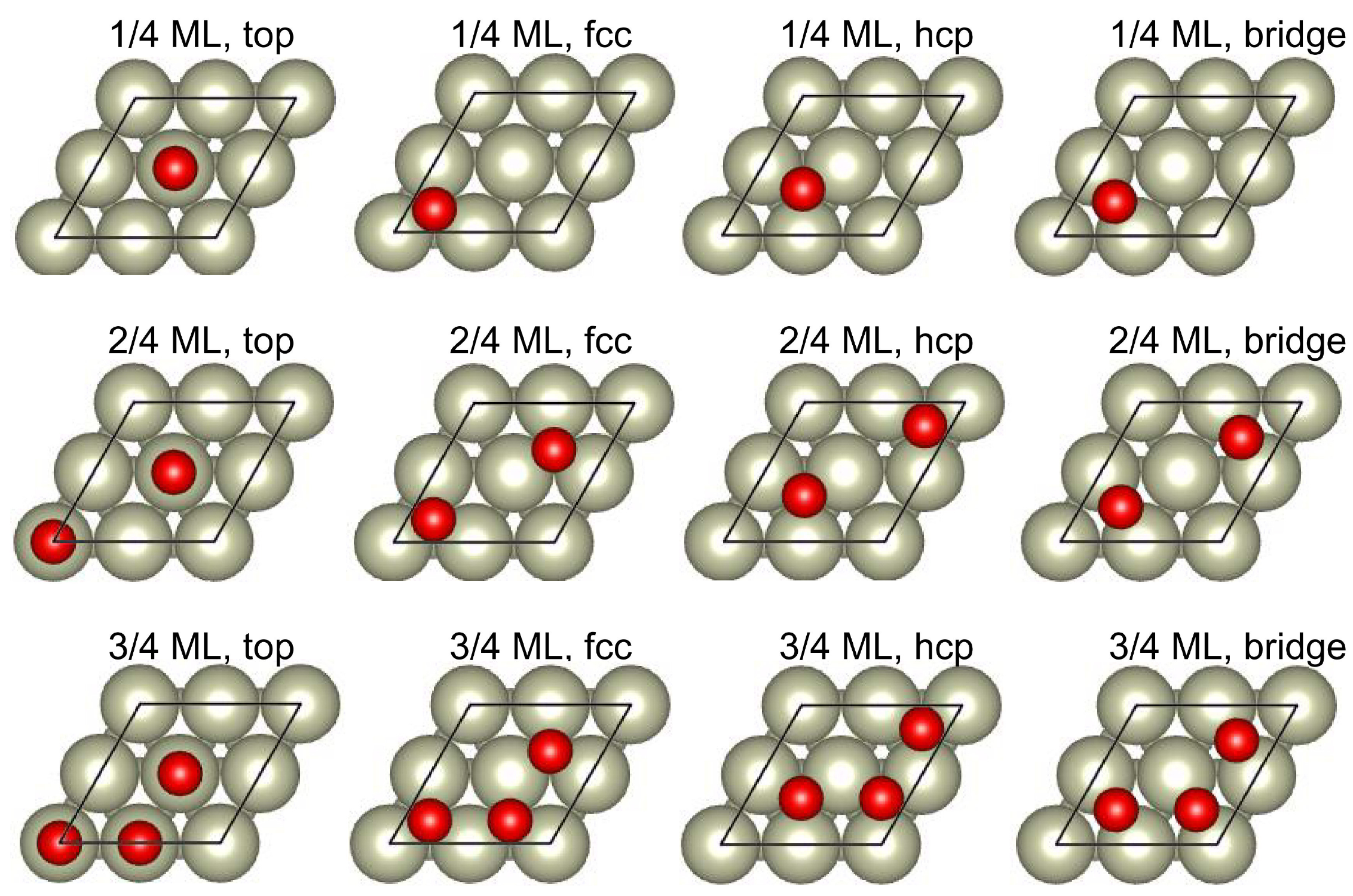}
\end{center}
\caption{Structural models used for calculating the adsorption energies of CO on the Rh(111) surface
in Table~\ref{tab:Eadsorption} and Table II of the main text.
The surfaces were modeled using a 2$\times$2 slab of 4 Rh layers with a vacuum width of 10 $\AA$.
The Rh, O, and C atoms are represented as light green, red, and brown balls, respectively.
The C atoms (in brown) are underneath the O atoms and thus not visible.
Note that most configurations considered here are not the ground-state structures
and these structures are not included in the training.
}
\label{fig:S22}
\end{figure}

\begin{table}[htbp]
\caption {The surface energy of the Rh(111) surface ($E_{\rm s}$) and
CO adsorption energies at different coverages ($E_{\rm ad}$, in eV per CO) predicted by different methods.
The results from the nonlocal vdW-DF functional of Dion \emph{et al.}~\cite{PhysRevLett.92.246401}
as well as the DFT-D3 method of Grimme \emph{et al.}~\cite{doi:10.1063/1.3382344}
using the revised PBE of Hammer \emph{et al.}~\cite{PhysRevB.59.7413} (RPBE+D3) are also given.
All calculations here were carried out using a 2$\times$2 slab of 4 Rh layers with a vacuum width of 10 $\AA$.
The adopted CO@Rh(111) structure models are displayed in Fig.~\ref{fig:S22}.
The RPA calculations were performed using the PBE relaxed structures.
}
\begin{ruledtabular}
\begin{tabular}{lccccccccc}
                           &     PBE  &    MTP-PBE & KMLFF-PBE   & vdW-DF & RPBE+D3 & RPA &  MTP-RPA$^\Delta$  &  KMLFF-RPA$^\Delta$ & Experiment \\
                           \hline
 $E_{\rm s}$ (eV/u.a.)                  &  0.813   &     0.809 &     0.805  &  0.780 & 1.103  &  1.027 &     1.020   &      1.020  & 1.039~\cite{TYSON1977267}\\
\\
$E_{\rm ad}$(1/4 ML, top)    & $-$1.916 &  $-$1.885 &  $-$1.879  & $-$1.663 & $-$1.994 &  $-$1.464 &  $-$1.482   &   $-$1.477 & 1.45$\pm$0.14~\cite{ABILDPEDERSEN20071747}\\
$E_{\rm ad}$(1/4 ML, fcc)    & $-$1.859 &  $-$1.861 &  $-$1.849  & $-$1.454 & $-$1.835 &  $-$1.221 &  $-$1.250   &   $-$1.249  &\\
$E_{\rm ad}$(1/4 ML, hcp)    & $-$1.950 &  $-$1.953 &  $-$1.964  & $-$1.530 & $-$1.917 &  $-$1.287 &  $-$1.322   &   $-$1.340  &\\
$E_{\rm ad}$(1/4 ML, bridge) & $-$1.842 &  $-$1.830 &  $-$1.819  & $-$1.478 & $-$1.856 &  $-$1.248 &  $-$1.280   &   $-$1.285  &\\
\\
$E_{\rm ad}$(2/4 ML, top)    & $-$1.610 &  $-$1.594 &  $-$1.580  & $-$1.389 & $-$1.693 &  $-$1.244 &  $-$1.247   &   $-$1.241  &\\
$E_{\rm ad}$(2/4 ML, fcc)    & $-$1.618 &  $-$1.607 &  $-$1.607  & $-$1.256 & $-$1.600 &  $-$1.085 &  $-$1.097   &   $-$1.108  &\\
$E_{\rm ad}$(2/4 ML, hcp)    & $-$1.670 &  $-$1.656 &  $-$1.669  & $-$1.296 & $-$1.646 &  $-$1.123 &  $-$1.139   &   $-$1.158  &\\
$E_{\rm ad}$(2/4 ML, bridge) & $-$1.739 &  $-$1.724 &  $-$1.709  & $-$1.410 & $-$1.763 &  $-$1.269 &  $-$1.271   &   $-$1.265  &\\
\\
$E_{\rm ad}$(3/4 ML, top)    & $-$1.323 &  $-$1.332 &  $-$1.302  & $-$1.136 & $-$1.421 &  $-$1.030 &  $-$1.056   &   $-$1.031  &\\
$E_{\rm ad}$(3/4 ML, fcc)    & $-$1.415 &  $-$1.404 &  $-$1.369  & $-$1.089 & $-$1.411 &  $-$0.961 &  $-$0.967   &   $-$0.975  &\\
$E_{\rm ad}$(3/4 ML, hcp)    & $-$1.446 &  $-$1.425 &  $-$1.413  & $-$1.109 & $-$1.437 &  $-$0.985 &  $-$0.982   &   $-$1.012  &\\
$E_{\rm ad}$(3/4 ML, bridge) & $-$1.404 &  $-$1.404 &  $-$1.316  & $-$1.106 & $-$1.414 &  $-$1.004 &  $-$0.967   &   $-$0.949  &\\
\end{tabular}
\end{ruledtabular}
\label{tab:Eadsorption}
\end{table}

\begin{table}[htbp]
\caption {The bond lengths (in $\AA$) of C-O and C-Rh for CO adsorbed on the top or hcp sites
predicted by different methods. Here, the surface is modeled using a 2$\times$2 slab of 4 Rh layers and a vacuum width of 10 $\AA$
and only one CO molecule is considered, leading to 1/4 ML coverage.
RPA and PBE predict almost identical C-O bond lengths for both the top and hcp sites, while RPA
predicted larger C-Rh bond lengths than PBE for both sites.
}
\begin{ruledtabular}
\begin{tabular}{lccccc}
& \multicolumn{2}{c}{top site} && \multicolumn{2}{c}{hcp site} \\
 \cline{2-3}  \cline{5-6}
                       &  $d_{\rm C-O}$  &$d_{\rm C-Rh}$  &&    $d_{\rm C-O}$   & $d_{\rm C-Rh}$ \\
\hline
PBE                    &     1.165       & 1.828      &&     1.197        &  2.085 \\
MTP-PBE                &     1.165       & 1.827      &&     1.196        &  2.081 \\
KMLFF-PBE              &     1.165       & 1.829      &&     1.196        &  2.082 \\
MTP-RPA$^\Delta$       &     1.164       & 1.857      &&     1.193        &  2.109 \\
KMLFF-RPA$^\Delta$     &     1.164       & 1.857      &&     1.193        &  2.106 \\
\end{tabular}
\end{ruledtabular}
\label{tab:bond_length}
\end{table}

\begin{figure}
\begin{center}
\includegraphics[width=0.80\textwidth, clip]{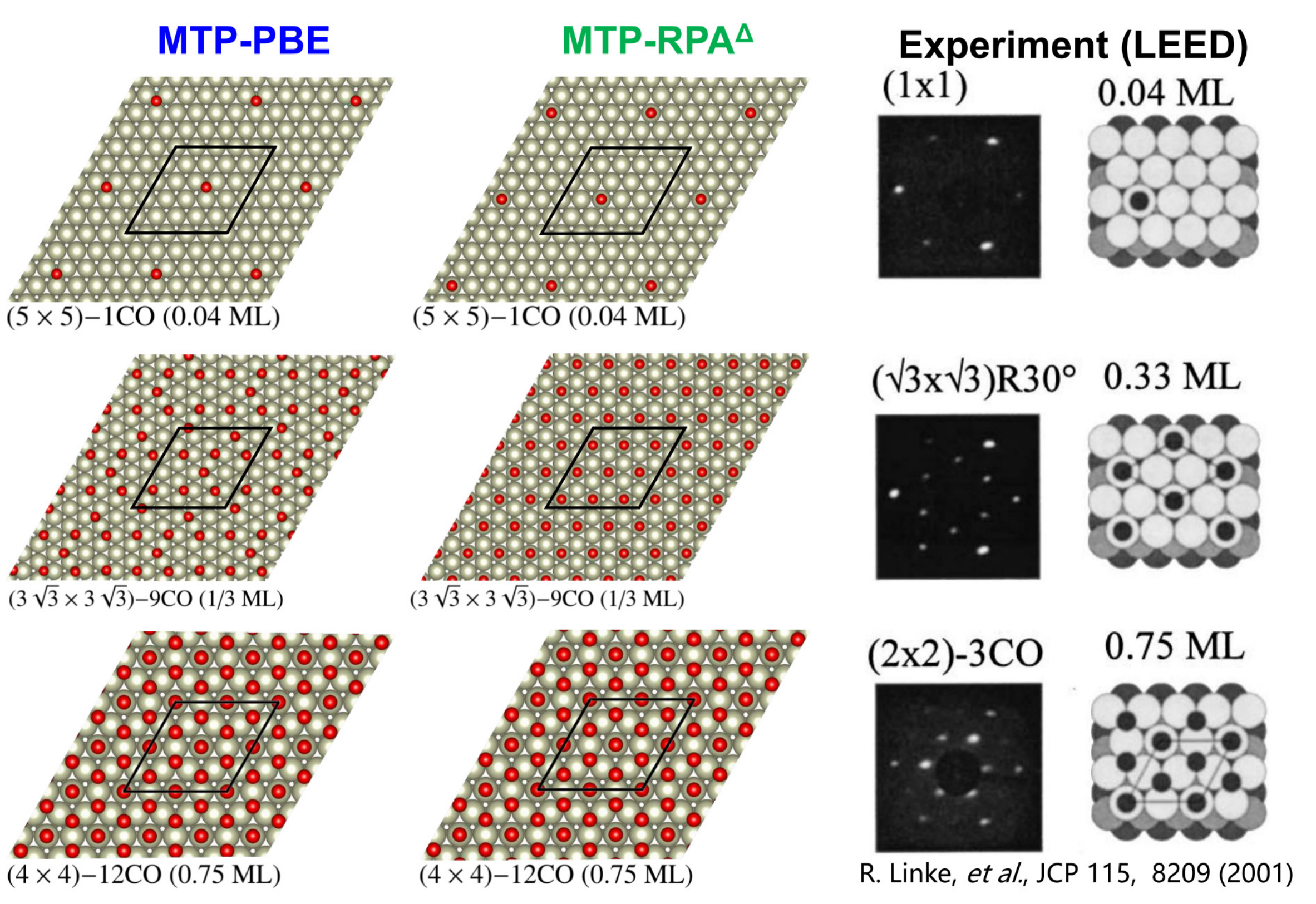}
\end{center}
\caption{Ground-state structures of CO adsorption on the Rh(111) surface for different coverages predicted by MTP-PBE (left column) and MTP-RPA$^\Delta$ (middle column),
which are compared to the experimental data (right column) taken from Ref.~\cite{doi:10.1063/1.1355767}.
The black lines indicate the Rh(111) surface that is periodically replicated.
}
\label{fig:theory_vs_Expt.}
\end{figure}

\section{Data availability}
The data that support the findings of this study are available at
{https://ucloud.univie.ac.at/index.php/s/A8vmHSHP44XbMY1}

\bibliographystyle{apsrev4-1}
\bibliography{reference} 